\documentclass[journal]{IEEEtran}
\usepackage[table]{xcolor}
\usepackage{cite}
\usepackage{graphicx}
\usepackage{lipsum}
\usepackage{stfloats}
\usepackage{bm}
\usepackage{amsmath,amssymb}
\usepackage{amsthm}
\usepackage{pifont}
\usepackage{bbding}
\usepackage{booktabs}
\usepackage{ctable}
\usepackage{threeparttable}
\usepackage{footmisc}
\usepackage{framed} 
\usepackage{array}
\usepackage{multirow}
\usepackage{longtable}
\usepackage{rotating}
\usepackage{fancyhdr}
\usepackage{lastpage}
\usepackage{layout}
\usepackage{wrapfig}
\usepackage{xcolor}
\usepackage{tabularx}
\usepackage{tabu}
\usepackage{enumitem}
\usepackage{url}
\usepackage{dsfont}
\usepackage{arydshln}
\usepackage{algorithm}
\usepackage{algorithmic}
\usepackage{makecell}
\usepackage{enumerate}

\makeatletter
\newcommand{\removelatexerror}{\let\@latex@error\@gobble}
\makeatother

\makeatletter
\newenvironment{breakablealgorithm}
  {
   \begin{center}
     \refstepcounter{algorithm}
     \hrule height.8pt depth0pt \kern2pt
     \renewcommand{\caption}[2][\relax]{
       {\raggedright\textbf{\ALG@name~\thealgorithm} ##2\par}%
       \ifx\relax##1\relax 
         \addcontentsline{loa}{algorithm}{\protect\numberline{\thealgorithm}##2}%
       \else 
         \addcontentsline{loa}{algorithm}{\protect\numberline{\thealgorithm}##1}%
       \fi
       \kern2pt\hrule\kern2pt
     }
  }{
     \kern2pt\hrule\relax
   \end{center}
  }
\makeatother

\newtheorem{myPropos}{\textbf{Proposition}}
\graphicspath{{Figures/}}
\DeclareGraphicsExtensions{.png, .eps}

\begin{document}
	\title{Joint Knowledge and Power Management for Secure Semantic Communication Networks}
	\author{
	Xuesong Liu,~\IEEEmembership{Student Member,~IEEE},
	Yansong Liu,~\IEEEmembership{Member,~IEEE},
				 Haoyu Tang,~\IEEEmembership{Member,~IEEE},\\
				 Fangzhou Zhao,~\IEEEmembership{Student Member,~IEEE},
				Le Xia,~\IEEEmembership{Member,~IEEE},
				 and Yao Sun,~\IEEEmembership{Senior~Member,~IEEE}
	\thanks{
	
	
	Xuesong Liu, Fangzhou Zhao, Le Xia, and Yao Sun are with the James Watt School of Engineering, University of Glasgow, Glasgow G12 8QQ, UK (e-mail: \{2928395L, 2430965z\}@student.gla.ac.uk; xiale1995@outlook.com,Yao.Sun@glasgow.ac.uk;).
	
	Yansong Liu (\textit{Corresponding author: Yansong Liu}.) is with the School of Intelligent Engineering, Shandong Management University, China 250357 (e-mail: lys@sdmu.edu.cn).
	
	Haoyu Tang is with the School of Software, Shandong University, China 250101 (e-mail: tanghao258@sdu.edu.cn).
%
	
}
}	
	\maketitle
	\begin{abstract}
	Recently, semantic communication (SemCom) has shown its great superiorities in resource savings and information exchanges.
	However, while its unique background knowledge guarantees accurate semantic reasoning and recovery, semantic information security-related concerns are introduced at the same time.
	Since the potential eavesdroppers may have the same background knowledge to accurately decrypt the private semantic information transmitted between legal SemCom users, this makes the knowledge management in SemCom networks rather challenging in joint consideration with the power control.
	To this end, this paper focuses on jointly addressing three core issues of power allocation, knowledge base caching (KBC), and device-to-device (D2D) user pairing (DUP) in secure SemCom networks.
	We first develop a novel performance metric, namely semantic secrecy throughput (SST), to quantify the information security level that can be achieved at each pair of D2D SemCom users.
	Next, an SST maximization problem is formulated subject to secure SemCom-related delay and reliability constraints.
	Afterward, we propose a security-aware resource management solution using the Lagrange primal-dual method and a two-stage method.
	Simulation results demonstrate our proposed solution nearly doubles the SST performance and realizes less than half of the queuing delay performance compared to different benchmarks.
	
	\end{abstract}
	

	\IEEEpeerreviewmaketitle
	
	\section{Introduction}
	\IEEEPARstart{S}{emantic} communication (SemCom) has been recognized as a transformative paradigm that can greatly enhance information exchange efficiency and robustness by shifting the delivery focus from bits to semantics~\cite{zhang2022toward}. By employing advanced deep learning (DL) algorithms, a semantic encoder is deployed at the transmitter of the SemCom system to extract core semantic features and filter out extraneous content from the source information, thus reducing the number of bits required. Correspondingly, the receiver embeds a semantic decoder, which is jointly trained with the semantic encoder, to precisely recover the meaning from the received bits, even in the presence of bit errors caused by signal distortions under harsh channel conditions~\cite{luo2022semantic}. Most importantly, the semantic encoder and decoder should have the equivalent background knowledge for semantic reasoning and interpretation, and the higher the knowledge matches, the more accurate the semantics can be recovered.

	The most common manner now to achieve SemCom is device-to-device (D2D) transmission~\cite{ye2018end}, in which the delivery of semantic information is further improved, especially in scenarios with limited bandwidth, network congestion, or device arithmetic constraints~\cite{r6,niazmand2025joint,zhuang2019sdn}.
 		Although D2D SemCom has great potential in improving communication efficiency and task execution, it is also facing new security and privacy challenges in coordination with the requirement of background knowledge matching among multiple SemCom users.
 		In recent studies, a key concept of knowledge base (KB) is introduced in SemCom, which is deemed a small entity containing a specific domain of knowledge that can be cached in advance in local devices or shared during communication~\cite{sun2024s}.
 		To be more concrete, the base station (BS) is a central knowledge scheduler responsible for distributing KBs to its serving users.
 		However, eavesdroppers equipped with advanced coding models and the same KBs may exist in the cellular area, aiming to eavesdrop private information from D2D SemCom links.
 		Especially if there exist some eavesdroppers adjacent to a D2D SemCom pair, the KBs shared between the two D2D users may be eavesdropped by malicious users and then the transmitted semantics can also be recovered by others, which may cause severe communication security issues.
 		In this case, traditional information security performance metrics, e.g., secrecy bit throughput or secrecy outage probability~\cite{du2023rethinking}, are no longer applicable to such a case, since these eavesdroppers concentrate more upon stealing semantic information rather than bit information.
 		Accordingly, there is a pressing demand to explore the protection of semantic information for D2D SemCom underlying cellular networks.
 		
 		In parallel, it is worth pointing out that different resource management solutions, particularly in knowledge and power allocation, can affect the overall semantic secrecy performance of D2D SemCom networks to a certain degree.
 		For instance, in a scenario with multiple SemCom users (SUs) and eavesdroppers, if each SU is assigned with the suitable KBs matching its paired SU, the semantic information throughput that can be conveyed by the corresponding D2D link is obviously greater than that can be eavesdropped.
 		This is due to the unique knowledge equivalence mechanism in SemCom, where the more the background knowledge of the transceiver matches, the more the semantic recovery rate can be accurate, i.e., more semantic information can be precisely obtained by the receiver.
 		Apart from this, the power control at each SU is paramount of importance, which has been a classical communication security challenge in cellular networks.
 		Generally, given the fixed relative position of two SUs and one eavesdropper, if the distance between two paired SUs is closer than that between the transmitter SU and eavesdropper, the higher the transmit power, the more the bit secrecy throughput, and vice versa.
 		Hence, both the power and KB management can lead to different semantic information secrecy performance of the SemCom network, which becomes rather critical and tricky, especially when jointly considering their optimality. 

		In fact, there has been some noteworthy related works paving the way for the development of SemCom.
		Benefit from advanced DL algorithms, the authors in \cite{r4} developed efficient deep multiple access in D2D scenarios to improve the performance of image transmission.
		\cite{xia2024generative} developed a generative AI-integrated end-to-end SemCom framework in a cloud-edge-mobile design for multimodal AIGC provisioning.
		In addition, a series of studies related to secure SemCom have also emerged.
		In \cite{r5}, the authors proposed a privacy-preserving SemCom framework that leverages knowledge to protect sensitive information.
		Meanwhile, in \cite{r6}, the authors pointed out the risk of eavesdropping on knowledge and identify resource allocation schemes in KBs as a viable future research direction.
		Some other works~\cite{du2023rethinking,shen2023secure,yang2024secure} also surveyed and discussed semantic information security in a comprehensive way, in particular comparing it to classical physical layer security.
		For the unique requirements of knowledge pairing in D2D SemCom, \cite{xia2023xurllc} proposed a reliable and low-latency knowledge matching method to solve the problem of semantic service provisioning among multiple vehicle-to-vehicle SemCom users.
		Besides, \cite{10122232} developed a long-term robust resource allocation strategy to satisfy the trade-off between user satisfaction, queue stability and communication latency for D2D SemCom.
	
		Nevertheless, to the best of our knowledge, none of these papers technically investigate the semantic information security-aware resource management problem for D2D SemCom.
		In this work, our main task is to seek optimal resource allocation in~\textit{secure SemCom network} (SSCN) to reach the best overall semantic network performance. Especially considering the unique knowledge matching demands between multiple SUs and the existence of eavesdroppers, devising the best resource management solution becomes quite indispensable, which can yield many benefits, such as further improving bandwidth utilization, and ensuring secure, efficient, and high-quality SemCom service provisioning.
		In view of these unique requirements, there exist three fundamental networking challenges as follows.
	\begin{itemize}
	\item \textit{Challenge 1: How to find an appropriate metric to measure semantic network performance related to secure SemCom?}
	The longer it takes to allocate knowledge, the more susceptible transmitted information is to eavesdropping. Considering this, scheduling the delay a semantic packet spends in the queue buffer of each D2D SemCom link is of paramount importance. Moreover, traditional system metrics, e.g., bit throughput, is no longer applicable to measure the SSCN. Especially taking into account the presence of eavesdroppers and semantic information importance in SemCom, how to define appropriate secure SemCom-related metrics should be the first challenge.
	\item \textit{Challenge 2: How to accurately assign KBs to multiple SemCom users with different knowledge preferences?}
	In SemCom, the KB is the key to perform semantic inference and recovery. Considering that each SemCom user has its own knowledge preference and KB storage capacity, it is quite tricky for each user to realize the best caching policy for size-varying KBs. In particular, potential eavesdroppers pose continuous threats to semantic transmission. If being aware the knowledge distribution at eavesdroppers, how to achieve a secure knowledge base caching (KBC) model should be the second difficulty.
	\item \textit{Challenge 3: How to determine the best D2D SemCom pairing strategy among all uses to maximize knowledge protection against eavesdropping?}
	Note that high semantic fidelity can be guaranteed based on equivalent knowledge matching conditions between two SemCom users.
	In line with different KBC situations, varying channel conditions, and the potential eavesdropping, it poses the third challenge to find the best D2D user pairing (DUP) for all SemCom users in order to realize the optimal semantic secrecy information delivery.
	\end{itemize}
	
	To address the above gaps that have never been jointly investigated in the current literature, in this paper, we jointly optimize KBC, DUP, and power allocation in the SSCN with the consideration of unique SemCom characteristics and semantic information security.
	Both theoretical analysis and numerical results validate the performance superiority of our proposed solution in terms of semantic secrecy throughput, queuing delay, and semantic knowledge satisfaction.
	In a nutshell, our main contributions are summarized in the following:
	\begin{itemize}
		\item We first identify two core resource management problems of KBC and DUP and construct their mathematical models. Specially, considering the potential eavesdropping center and its knowledge distribution conditions, we develop a novel metric, namely semantic secrecy throughput (SST), to measure the overall semantic information security for all SemCom users.
		\item By taking into account personal preference for different knowledge bases, we derive the average queuing latency and define the average semantic knowledge satisfaction for each potential D2D SemCom link in a mathematical manner. In this way, a joint SST-maximization problem is then formulated for power allocation, KBC, and DUP subject to several practical constraints.
		\item We propose an optimal security-aware resource management solution with polynomial-time complexity. Specifically, a Lagrange primal-dual method is first utilized to obtain the dual problem, which is able to be decomposed to multiple subproblems. Without loss of optimality, in each iteration, we leverage a two-stage method to separately solve these subproblems, where the first stage is to determine the KBC and power allocation strategies and the second stage is to finalize the DUP policy.
	\end{itemize}
	
	The remainder of this paper is organized as follows.
	Section II first introduces the system model of SSCN and formulates a joint security-aware resource optimization problem.
	Then, we present our proposed solution and complexity analysis in Section III.
	Numerical results are demonstrated and discussed in Section IV, followed by the conclusions in Section V.

    \section{System Model and Problem Formulation}
	In this section, the considered SSCN scenario is first elaborated along with the knowledge caching model, secrecy queuing model, and semantic secrecy throughput model. Then, the corresponding optimization problem is formulated.
	\subsection{SSCN Scenario}
	\begin{figure}[t]
		\centering
		\includegraphics[width=0.48\textwidth]{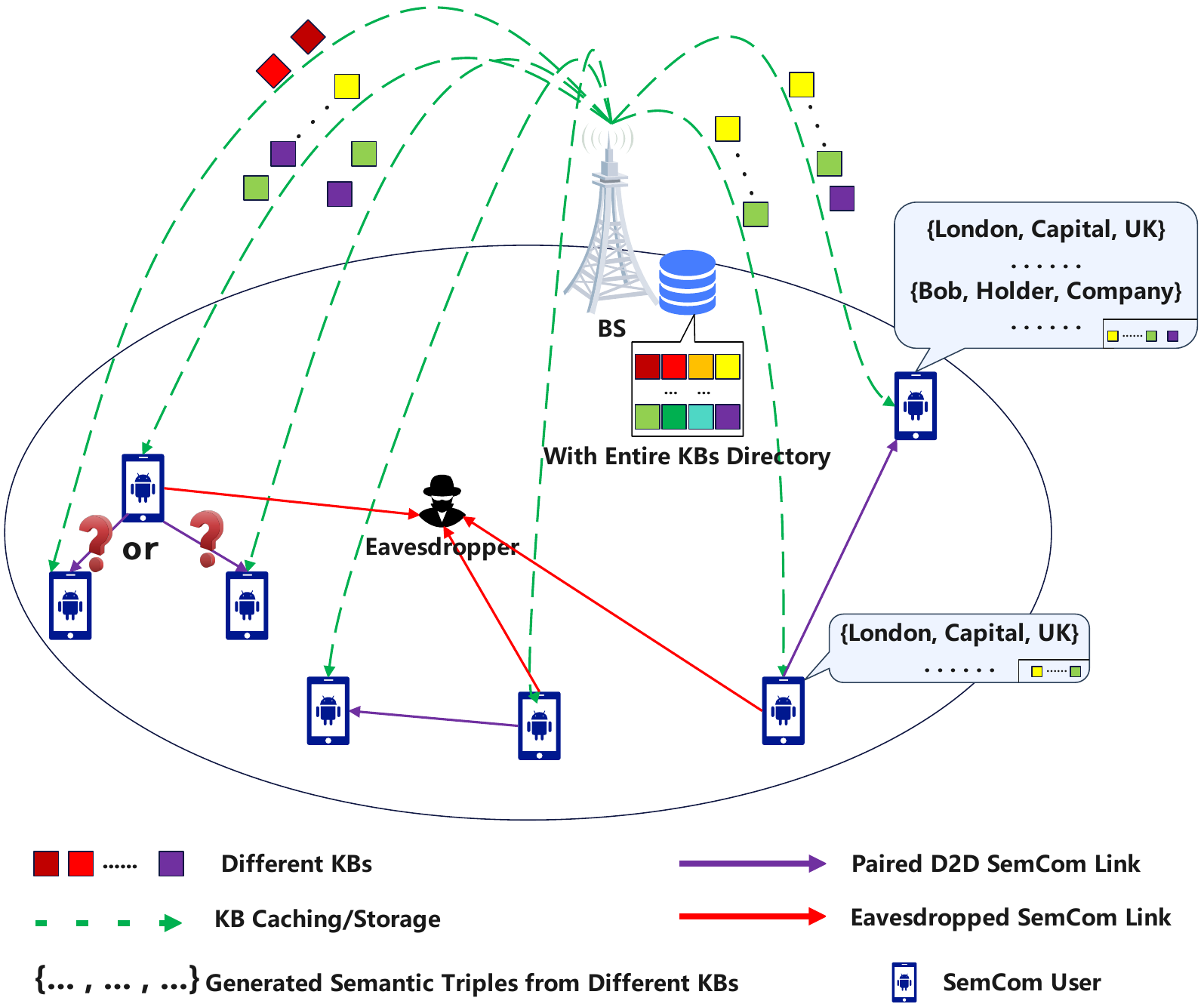} 
		\caption{The SSCN with KBC, DUP, and PC.}
		\label{Scenario}
    \end{figure}
	Consider a single-cell SSCN scenario as shown in Fig.~\ref{Scenario}, a total of $M$ SUs are distributed within the coverage of one BS, and each SU $i \in \mathcal{M}=\{1,2,\cdots,M\}$ is capable of providing D2D SemCom services to others.
	Without loss of generality, each SU is stipulated to be paired with only one (another) SU at a time for D2D SemCom.
	Herein, let $\beta_{i,j} \in \{0,1\}$ denote the binary DUP indicator for a potential SU $i$-SU $j$ ($j \in \mathcal{M}, j \neq i$) pair, where $\beta_{i,j}=1$ means SU $i$ is paired with SU $j$ for D2D SemCom, otherwise $\beta_{i,j}=1$.
	Besides, each D2D SemCom link has been pre-allocated an orthogonal subchannel with equal channel bandwidth $W$, and each SU $i$ has a maximum allowable transmit power $P_{\mathit{max}}$ for SemCom.
	Meanwhile, assume that there exists an eavesdropping center with a fixed spatial location, constantly attempting to eavesdrop on the conveyed semantic information from all D2D SemCom links and using the same subchannel and bandwidth resources as the correspondingly target links.
	Having these, let $G_{i,j}^{D}$ denote the channel power gain of the D2D link between the transmitter SU $i$ and the receiver SU $j$, and let $G_{i}^{E}$ denote the channel power gain of the eavesdropping link between the transmitter SU $i$ and the eavesdropping center.
	Further denoting the transmit power of SU $i$ as $P_{i}$, the signal-to-noise ratio (SNR) experienced by the D2D link is $\gamma_{i,j}^{D} = P_{i}G_{i,j}^{D}/\zeta^{2}$, where $\zeta^{2}$ is the noise power.
	Likewise, the SNR of the eavesdropping link regarding SU $i$ is $\gamma_{i}^{E} = P_{i}G_{i}^{E}/\zeta^{2}$.
	Hence, the bit rates at the D2D link and the eavesdropping link should be $r_{i,j}^{D} = W\log_{2}\left(1+\gamma_{i,j}^{D}\right)$ and $r_{i}^{E} = W\log_{2}\left(1+\gamma_{i}^{E}\right)$, respectively.
	
	Note that we allow SU $i$ to provide D2D SemCom services to SU $j$ only if the SNR $\gamma_{i,j}$ is above a specified threshold $\gamma_{0}$, hence the set of eligible communication neighbors of SU $i$ is defined as $\mathcal{M}_{i}=\{j \mid j \in \mathcal{M}, j \neq i, \gamma_{i,j} \geqslant \gamma_{0}\}$.
	Moreover, let the BS act as a semantic service controller to efficient schedule and coordinate the whole secure D2D SemCom process based on the request and state information received from all participating SUs within its coverage.
	
	\subsection{Knowledge Base Caching Model}
	The acquisition of necessary background knowledge is known to be inevitable for accurate semantic inference and interpretation in SemCom.
	Note that each KB contains the background knowledge of only one particular application domain (e.g., music or sports), and thus different KBs are associated with different background knowledge, and holding some common KBs becomes the necessary condition to perform SemCom between two D2D SUs in accordance with the knowledge equivalence principle.\footnote{The structure of a KB can roughly cover multiple computational ontologies, facts, rules and constraints associated to a specific domain~\cite{chein2008graph}.
    In recent deep learning-driven semantic coding models, the KB is also treated as a training database serving a certain class of learning tasks~\cite{luo2022semantic}.}
	Especially considering the existence of the eavesdropping center, different knowledge base caching (KBC) schemes can significantly affect the overlapping degree of background knowledge between different nodes, since the transmitted semantic information can only be decoded under the equivalent background knowledge.
	In this work, suppose that there is a KB library $\mathcal{K}$ with a total of $K$ differing KBs in the considered SSCN, and each KB $k \in \mathcal{K}=\left\{1, 2,\ldots, K\right\}$ is corresponding to a specific type of SemCom service.
	On this basis, all SUs should proactively download and cache their respective interested KBs from the BS to achieve D2D SemCom based on the desired semantic services.
	Note that each requires a unique storage size $s_{k}$, and each SU $i\in \mathcal{M}$ has a finite capacity $C_{i}$ for its local KB storage.
	In this way, we define a binary KBC indicator as
	\begin{equation}
		\label{alpha}
			\alpha_{i}^{k}=\left\{\begin{aligned}
			1,\quad &  \text{if KB $k$ is cached on SU $i$;}\\
			0,\quad &  \text{otherwise.}
		\end{aligned}
		\right.
	\end{equation}
	It is worth mentioning that the same KB cannot be cached repeatedly at one SU for reducing redundancy and for promoting the storage efficiency.
	
	Besides, it is noticed that different SUs may have different personal preferences for these KBs, thus resulting in the diversity of KB popularity.
	Naturally, the more popular the KBs, the higher the KBC probabilities.
	Herein, we assume that the KB popularity at each SU follows the Zipf distribution~\cite{piantadosi2014zipf,xia2023knowledge}.\footnote{Note that other known probability distributions can also be adopted for KB popularity without changing the remaining modeling and solution.}
	Hence, the probability of SU $i$ requesting its desired KB $k$-based SemCom services (generating the corresponding semantic data packets) is $p_{i}^{k}=\left(r_{i}^{k}\right)^{-\xi_{i}}/\sum_{e \in \mathcal{K}}e^{-\xi_{i}}, \forall (i,k) \in \mathcal{M}\times \mathcal{K}$, where $\xi_{i}$ ($\xi_{i}\geqslant 0$) is the skewness of the Zipf distribution, and $r_{i}^{k}$ is the popularity rank of KB $k$ at SU $i$.\footnote{The KB popularity ranking of each SU can be analyzed and estimated based on its historical messaging records~\cite{hassine2015popularity,xia2023xurllc,li2018service}, which will not be discussed in this paper.}
	Based on $p_{i}^{k}$, we specially define a KBC-related metric $\eta_{i}$, namely~\textit{semantic knowledge satisfaction}, to measure the satisfaction degree of SU $i$ caching its interested KBs, given as $\eta_{i}=\sum_{k \in \mathcal{K}}\alpha_{i}^{k}p_{i}^{k}$.
	It is further noticed that $\eta_{i} \geqslant \eta_{0}$, where $\eta_{0}$ is the unified minimum threshold that needs to be achieved at each SU.
	
	As for the eavesdropping center, let us assume it has its own KB preference for illegal eavesdropping and decoding of their interested semantic information from these D2D SemCom links.
	Without loss of the generality, let $r_{E}^{k}$ and $\xi_{E}$ ($\xi_{E}\geqslant 0$) denote the popularity rank of KB $k$ at the eavesdropping center and the skewness of its Zipf distribution, respectively.
	As such, the probability of KB $k$ being held by the eavesdropping center can be expressed by $p_{E}^{k}=\left(r_{E}^{k}\right)^{-\xi_{E}}/\sum_{e \in \mathcal{K}}e^{-\xi_{E}}, \forall k \in \mathcal{K}$.
	In other words, if denoting $\alpha_{E}^{k}$ as the KBC indicator of the eavesdropping center, where $\alpha_{E}^{k}=1$ indicates that KB $k$ is cached and $\alpha_{E}^{k}=0$ otherwise, we have $\Pr\left\{\alpha_{E}^{k}=1\right\}=p_{E}^{k}$ and $\Pr\left\{\alpha_{E}^{k}=0\right\}=1-p_{E}^{k}$.
%
	
	\subsection{Secrecy Delay Model}
	In secure communications, the longer the delay incurred in the communication process, the more likely it is that the transmitted information will be eavesdropped on~\cite{li2021secure}.
	Being aware of this, the knowledge matching based semantic packet queuing delay is jointly considered as a long-term metric in this work, which is to characterize the average sojourn time of semantic data packets in the receiver SU's queue buffer (following the first-come first-serve rule).
	This is mainly considered from the large-timescale perspective, as we only focus on the queuing latency under the steady-state of the semantic packet queuing system.
	Note that each semantic data packet is associated with a specific service type, i.e., a certain KB, while we assume that semantic data packets generated based on different KBs can co-exist in the queue and have independent average arrival rate and interpretation time.
	Besides, not all semantic data packets arriving at the receiver SU are always allowed to enter its queue, as some of them may mismatch the KBs currently held, rendering these packets uninterpretable~\cite{xia2023xurllc}.
	As for these mismatched packets, we can choose the traditional bit transmission, where the additional delay introduced can be easily covered by applying SemCom due to the considerably saved bits and communication delay.
	
	To preserve generality, we first suppose a Poisson data arrival process with average rate $\lambda_{i,j}^{k}=\lambda_{i,j}^{D}p_{i}^{k}$ for the D2D SemCom link between transmitter SU $i$ and receiver SU $j$ to account for semantic packet generation based on KB $k$, where $\lambda_{i,j}^{D}$ is the total arrival rate of all semantic packets from transmitter SU $i$ to receiver SU $j$.
	Herein, if assuming that each semantic packet has an average size of $L$ bits, given its achievable bit rate $r_{i,j}^{D}$ with variable $P_{i}$, we can approximately calculate $\lambda_{i,j}^{D}$ by $\lambda_{i,j}^{D}=r_{i,j}^{D}/L$.
	Combined with the KBC situation at SU $i$ and SU $j$, the effective arrival rate of semantic packets (i.e., semantic packets based on matched KB between them) in the queue is given as $\lambda_{i, j}^{\mathit{eff}}=\sum_{k \in \mathcal{K}}\alpha_{i}^{k}\alpha_{j}^{k}\lambda_{i,j}^{k}$.
	In parallel, let a random variable $I_{j}^{k}$ denote the Markovian interpretation time~\cite{lavee2009understanding} required by KB $k$-based packets at SU $j$ with mean $1/\mu_{j}^{k}$, which is determined by the computing capability of the SU and the type of the desired KB.
	However, since multiple packets based on different KBs are allowed to queue at the same time, it is seen that the interpretation time distribution for a receiver SU should be treated as a general distribution~\cite{ross2014introduction}.
	If further taking into account the KB popularity, we can calculate the ratio of the amount of KB $k$-based packets to the total packets in the SU $i$-SU $j$ pair's queue by $\epsilon_{i, j}^{k}=p_{i}^{k}/\sum_{f \in \mathcal{K}}\alpha_{i}^{f}\alpha_{j}^{f}p_{i}^{f}$.
	With the independence among packets based on different KBs, the interpretation time required by each packet in the queue is now expressed as $W_{i, j}=\sum_{k \in \mathcal{K}}\alpha_{i}^{k}\alpha_{j}^{k}\epsilon_{i, j}^{k}I_{j}^{k}$.
	
	Since the Markovian arrival process leads to the correlated packet arrivals while the service pattern of packets obeys a general distribution, the queue of each D2D pair in the SSCN can be modeled as an M/G/1 system, which has been widely used to model data traffic in wireless networks.
	According to the~\textit{Pollaczek-Khintchine formula}~\cite{pollaczek1930aufgabe}, the average queuing latency for the SU $i$-SU $j$ pair, denoted as $\delta_{i, j}$, is determined as follows\footnote{In order to guarantee the steady-state of the queuing system, a condition of $\lambda_{i, j}^{\mathit{eff}}\mathds{E}\left[W_{i, j}\right]<1$ must be satisfied before proceeding~\cite{ross2014introduction}. In this work, we assume that the packet interpretation rate is larger than the packet arrival rate to make the queuing latency finite and thus solvable.}
	\begin{equation}
		\delta_{i, j}=\frac{\lambda_{i, j}^{\mathit{eff}}\cdot \left(\mathds{E}^{2}\left[W_{i, j}\right]+\mathit{Var}\left(W_{i, j}\right)\right)}{2\left(1-\lambda_{i, j}^{\mathit{eff}}\cdot \mathds{E}\left[W_{i, j}\right]\right)}.\label{PK}
	\end{equation}
	To calculate the close form of $\delta_{i, j}$, again leveraging the independence of $I_{j}^{k}$ over $k$, we can obtain the expectation of the interpretation time for all semantic data packets by
	\begin{equation}
		\begin{aligned}
		\mathds{E}\left[W_{i, j}\right]=\sum_{k \in \mathcal{K}}\alpha_{i}^{k}\alpha_{j}^{k}\epsilon_{i, j}^{k}\mathds{E}\left[I_{j}^{k}\right]=\sum_{k \in \mathcal{K}}\frac{\alpha_{i}^{k}\alpha_{j}^{k}\epsilon_{i, j}^{k}}{\mu_{j}^{k}},\label{PK1}
		\end{aligned}
	\end{equation}
	and the variance of $W_{i, j}$ can be calculated by
	\begin{equation}
		\begin{aligned}
		\mathit{Var}\left(W_{i, j}\right)&=\sum_{k \in \mathcal{K}}\alpha_{i}^{k}\alpha_{j}^{k}\left(\epsilon_{i, j}^{k}\right)^{2}\mathit{Var}\left[I_{j}^{k}\right]\\
		&=\sum_{k \in \mathcal{K}}\alpha_{i}^{k}\alpha_{j}^{k}\left(\frac{\epsilon_{i, j}^{k}}{\mu_{j}^{k}}\right)^{2}.\label{PK2}
		\end{aligned}
	\end{equation}
	By substituting~(\ref{PK1}) and (\ref{PK2}) into (\ref{PK}), $\delta_{i, j}$ can be rewritten in~(\ref{Qdelay}), as shown at the bottom of the next page. Due to the security consideration, let $\delta_{0}$ denote a unified secrecy latency threshold that should be satisfied at all associated D2D SemCom links.
	\begin{figure*}[hb]
		\centering
		\hrulefill
		\begin{equation}
			\delta_{i, j}=\frac{\left[\left(\sum_{k \in \mathcal{K}}\alpha_{i}^{k}\alpha_{j}^{k}\frac{p_{i}^{k}/\mu_{j}^{k}}{\sum_{f \in \mathcal{K}}\alpha_{i}^{f}\alpha_{j}^{f}p_{i}^{f}}\right)^{2}+\sum_{k \in \mathcal{K}}\alpha_{i}^{k}\alpha_{j}^{k}\left(\frac{p_{i}^{k}/\mu_{j}^{k}}{\sum_{f \in \mathcal{K}}\alpha_{i}^{f}\alpha_{j}^{f}p_{i}^{f}}\right)^{2}\right]\cdot \left(\sum_{k \in \mathcal{K}}\alpha_{i}^{k}\alpha_{j}^{k}\lambda_{i,j}^{k}\right)}{2\left[1-\left(\sum_{k \in \mathcal{K}}\alpha_{i}^{k}\alpha_{j}^{k}\lambda_{i,j}^{k}\right)\cdot \left(\sum_{k \in \mathcal{K}}\alpha_{i}^{k}\alpha_{j}^{k}\frac{p_{i}^{k}/\mu_{j}^{k}}{\sum_{f \in \mathcal{K}}\alpha_{i}^{f}\alpha_{j}^{f}p_{i}^{f}}\right)\right]}.\label{Qdelay}
		\end{equation}
	\end{figure*}
	
	\subsection{Semantic Secrecy Throughput Measurement}
	Lately, a concept of \textit{semantic triplet} is dedicatedly introduced in the realm of SemCom to represent the interpretable relationship between two specific semantic entities implied in source information~\cite{yang2019transms,chen2020review,gao2024importance}, and its typical expression is~\textit{(Entity-A, Relationship, Entity-B)}, as depicted in the exemplification of Fig.~\ref{Scenario}.
	In the context of the above secrecy queuing latency model, we assume that each semantic data packet contains exactly one semantic triple, which is corresponding to a certain KB.
	To be more specific, all source information of transmitters in SemCom is first encoded in the minimum unit of semantic triplets by semantic encoding models.\footnote{This assumption is justified since core semantic information can be extracted by state-of-the-art DL models from multimedia services (e.g., text~\cite{xie2021deep}, image~\cite{shi2021semantic}, and video~\cite{xia2023wiservr}) to draw the semantic knowledge graph, which can be decomposed into multiple semantic triplets. In other words, it is reasonable that source information in any format can be conveyed in units of semantic triplets for SemCom service provisioning.}
	Afterward, each semantic triplet is encoded by each SU's channel encoder and then encapsulated into an $L$ bits-size semantic data packet for wireless transmission~\cite{gao2024importance}.
	As such, we clearly have the total number of semantic triplets transmitted from transmitter SU $i$ to receiver SU $j$ per second given by $\lambda_{i,j}^{D}$, as aforementioned, and likewise, obtain that for the eavesdropping link as $\lambda_{i}^{E}=r_{i}^{E}/L$.
	
	Since each KB has its popularity rank at each SU as identified in the KBC model, it means that each semantic triplet also has the same preference level.
	Naturally, transmitting the higher-ranked semantic triplets contributes more valuable semantic information for each SemCom receiver.
	Inspired by this, we employ a performance metric called~\textit{semantic value} proposed in~\cite{gao2024importance} to measure the semantic information importance of semantic triplets with different rankings at each SU.
	Proceeding as in~\cite{gao2024importance}, the calculation of semantic value for each KB $k$-based semantic triplet relies on SU $i$'s Zipf distribution, which is exactly the numerator of $p_{i}^{k}$, given by $\left(r_{i}^{k}\right)^{-\xi_{i}}$.
	Note that the semantic value above is measured according to the transmitter's KB preference only, which is reasonable as the primary purpose of SemCom is generally determined by the sender, rather than the receiver.
	With these and further combining the effective semantic packet arrival of KB $k$ from SU $i$ to SU $j$ given in $\lambda_{i, j}^{\mathit{eff}}$, i.e., $\alpha_{i}^{k}\alpha_{j}^{k}\lambda_{i,j}^{k}$, we can obtain the effective semantic value transmitted via the SU $i$-SU $j$ D2D pair per second as
	\begin{equation}
		V_{i,j}^{D} = \sum_{k\in \mathcal{K}}\alpha_{i}^{k}\alpha_{j}^{k}\lambda_{i,j}^{k}\left(r_{i}^{k}\right)^{-\xi_{i}}=\frac{r_{i,j}^{D}}{L}\sum_{k\in \mathcal{K}}\alpha_{i}^{k}\alpha_{j}^{k}p_{i}^{k}\left(r_{i}^{k}\right)^{-\xi_{i}}.\label{saww}
	\end{equation}
	Similarly, based on the semantic triplet transmission rate $\lambda_{i}^{E}$ and the KBC situation $\alpha_{E}^{k}$ pertinent to the eavesdropping center, the average number of KB $k$-based semantic triplets that can be eavesdropped per second should be
	\begin{equation}
		\begin{aligned}
		V_{i}^{E} &= \sum_{k\in \mathcal{K}}\lambda_{i}^{E}\alpha_{i}^{k}p_{i}^{k}\Pr\left\{\alpha_{E}^{k}=1\right\}\left(r_{i}^{k}\right)^{-\xi_{i}}\\
		&=\frac{r_{i}^{E}}{L}\sum_{k\in \mathcal{K}}\alpha_{i}^{k}p_{i}^{k}p_{E}^{k}\left(r_{i}^{k}\right)^{-\xi_{i}}.\label{saww2}
	\end{aligned}
	\end{equation}
	Therefore, drawing on the classical metric of secrecy throughput in traditional secure communications~\cite{du2023rethinking,qiao2011secure,ng2015multiobjective}, we define here a new metric, namely \textit{semantic secrecy throughput} (SST), by subtracting $V_{i}^{E}$ in \eqref{saww2} from $V_{i,j}^{D}$ in \eqref{saww} as
	\begin{equation}
		\begin{aligned}
			V_{i,j}^{S} &= \left[V_{i,j}^{D}-V_{i}^{E}\right]^{+}\\
			&=\frac{1}{L}\left[\sum_{k\in \mathcal{K}}\alpha_{i}^{k}\left(r_{i}^{k}\right)^{-\xi_{i}}\left(r_{i,j}^{D}\alpha_{j}^{k}p_{i}^{k}-r_{i}^{E}p_{i}^{k}p_{E}^{k}\right)\right]^{+},
		\end{aligned}
	\end{equation}
	where the operator $\left[\cdot\right]^{+}$ is to output the maximum value between its argument and zero.
	To provide an intuitive explanation, $V_{i,j}^{S}$ is the secure semantic value transmission rate at which the legitimate SU $i$-SU $j$ pair is able to communicate safely without enough valuable semantic information being stolen by the eavesdropping center.
	If $V_{i,j}^{D} > V_{i}^{E}$, the D2D SemCom users can transmit semantic information to each other with a secure rate of $(V_{i,j}^{D} - V_{i}^{E})$.
	Otherwise, any transmitted semantics at the SU $i$-SU $j$ pair will be stolen by the eavesdroppers.
	If taking into account all DUP possibilities, the overall SST of SSCN should be the sum of $\beta_{i,j}V_{i,j}^{S}$ corresponding to all potential SU $i$-SU $j$ cases, that is,
	\begin{equation}
	\label{14124}
		\begin{aligned}
		&\mathit{SST} = \sum_{i \in \mathcal{M}}\sum_{j \in \mathcal{M}_{i}}\beta_{i,j}V_{i,j}^{S}\\
		&=\sum_{i \in \mathcal{M}}\sum_{j \in \mathcal{M}_{i}}\frac{\beta_{i,j}}{L}\left[\sum_{k\in \mathcal{K}}\alpha_{i}^{k}\left(r_{i}^{k}\right)^{-\xi_{i}}\left(r_{i,j}^{D}\alpha_{j}^{k}p_{i}^{k}-r_{i}^{E}p_{i}^{k}p_{E}^{k}\right)\right]^{+}.
		\end{aligned}
	\end{equation}
 	
	\subsection{Problem Formulation}
	For ease of illustration, we first define three variable sets $\bm{\alpha}=\left\{\alpha_{i}^{k}\mid i \in \mathcal{M}, k \in \mathcal{K}\right\}$, $\bm{\beta}=\left\{\beta_{i,j}\mid i \in \mathcal{M}, j \in \mathcal{M}_{i}\right\}$, and $\bm{P}=\left\{P_{i}\mid i \in \mathcal{M}\right\}$ that consist of all possible indicators pertinent to KBC, DUP, and PC, respectively.
	Without loss of generality, the objective of our optimization is to maximize $\mathit{SST}$ in \eqref{14124} by jointly optimizing $(\bm{\alpha},\bm{\beta}, \bm{P})$, while subject to SemCom-relevant latency and performance requirements alongside several practical system constraints.
	The problem is specifically formulated as follows:
	\begin{align}
	\mathbf{P0}:\ \max_{\bm{\alpha},\bm{\beta}, \bm{P}} \quad & \sum_{i \in \mathcal{M}}\sum_{j \in \mathcal{M}_{i}}\beta_{i,j}V_{i,j}^{S}~\label{P0}\\
	{\rm s.t.} \quad &  \sum_{k \in \mathcal{K}} \alpha_{i}^{k}\cdot s_{k}\leqslant C_{i},\  \forall i\in \mathcal{M},\tag{\ref{P0}a}\\
	& \eta_{i}\geqslant \eta_{0},\ \forall i \in \mathcal{M},\tag{\ref{P0}b}\\
	&\sum_{j \in \mathcal{M}_{i}}\beta_{i, j}= 1,\ \forall i \in \mathcal{M},\tag{\ref{P0}c}\\
	& \beta_{i, j}=\beta_{j, i},\ \forall \left( i,j\right) \in \mathcal{M}\times \mathcal{M}_{i},\tag{\ref{P0}d}\\
	& \sum_{j \in \mathcal{M}_{i}}\beta_{i, j}\delta_{i, j}\leqslant \delta_{0},\ \forall i \in \mathcal{M},\tag{\ref{P0}e}\\
	& \sum_{j \in \mathcal{M}_{i}}\beta_{i, j}V_{i,j}^{S}\geqslant V_{0},\ \forall i \in \mathcal{M},\tag{\ref{P0}f}\\
	& \alpha_{i}^{k}\in \{0,1\},\ \forall \left( i,k\right) \in\mathcal{M}\times \mathcal{K},\tag{\ref{P0}g}\\
	& \beta_{i,j}\in \{0,1\}, \ \forall \left( i,j\right) \in\mathcal{M}\times \mathcal{M}_{i}\tag{\ref{P0}h}\\
	& 0\leqslant P_{i}\leqslant P_{max},\ \forall i\in\mathcal{M}.\tag{\ref{P0}i}
	\end{align}
	Constraints (\ref{P0}a) guarantees that the total size of KBs cached at each SU cannot exceed its maximum storage capacity, while constraint (\ref{P0}b) represents the aforementioned semantic knowledge satisfaction requirement.
	Then, constraints (\ref{P0}c) and (\ref{P0}d) jointly model the single-D2D pairing limitation of SUs.
	Next, constraint (\ref{P0}e) indicates the maximum secrecy queuing latency threshold, while constraint (\ref{P0}f) is the minimum SST requirement at each D2D SemCom link indicating the reliability level of the system.
	Moreover, constraints (\ref{P0}g) and (\ref{P0}h) characterize the binary properties of $\bm{\alpha}$ and $\bm{\beta}$.
	Finally, constraint (\ref{P0}i) determines the transmit power range for all SUs.

	Carefully examining $\mathbf{P0}$, it can be observed that the optimization is quite challenging to be solved straightforwardly due to several intractable mathematical obstacles.
	First of all, $\mathbf{P0}$ is an NP-hard optimization problem as demonstrated below.
	Consider a special case of $\mathbf{P0}$ where all $\bm{\beta}$- and $\bm{P}$-related constraints have been satisfied.
    In this case, constraints (\ref{P0}c)-(\ref{P0}d) and (\ref{P0}h)-(\ref{P0}i) can all be removed, and the primal problem degenerates into a variant $0$-$1$ multi-knapsack problem that is known to be NP-hard~\cite{kellerer2004multidimensional}, thereby $\mathbf{P0}$ is also NP-hard.
	Besides, $\mathbf{P0}$ involves both continuous and discrete variables, and its objective function (\ref{P0}) is quite complicated alongside constraints (\ref{P0}e) and (\ref{P0}f), which prevents us from using the conventional two-step solution (i.e., relaxation and recovery) to approach optimality.
	In more detail, the problem after relaxing $\bm{\alpha}$ and $\bm{\beta}$ should still be a nonconvex optimization problem owing to the non-convexity preserved in (\ref{P0}) and constraints (\ref{P0}e) and (\ref{P0}f).
	Therefore, a severe performance penalty and a high computational complexity will be incurred from the procedure of integer recovery due to the huge performance compromise on solving the nonconvex problem for relaxed variables~\cite{tuy1998convex,papadimitriou1998combinatorial,burer2012non}.
	In view of the above mathematical challenges, we propose an efficient resource allocation strategy in the next section to reach the optimality of $\mathbf{P0}$ and obtain the joint KBC, DUP, and PC solution.
	
	\section{Proposed Security-Aware Resource Allocation for SSCN}
	In this section, we illustrate how to achieve the optimal security-aware resource management in the SSCN.
	We first employ the Lagrange dual method to transform the primal problem $\mathbf{P0}$ to its dual optimization problem.
	Then, given dual variables in each iteration, the dual problem is decomposed into multiple subproblems, which are solved by the proposed two-stage method.
	Finally, the workﬂow of our solution and its complexity analysis are provided.
	
	\subsection{Primal-Dual Problem Transformation}
	To make $\mathbf{P0}$ tractable, we first incorporate constraints (\ref{P0}e) and (\ref{P0}f) into the objective function (\ref{P0}) by associating two Lagrange multipliers $\bm{\tau}=\{\tau_{i}\mid i \in \mathcal{M}\}$ and $\bm{\rho}=\{\rho_{i}\mid i\in \mathcal{M}\}$.
	As such, its Lagrange function is found by \eqref{Lagrangian}, as shown at the bottom of the next page, in which $\widetilde{L}_{\bm{\tau},\bm{\rho}}\left(\bm{\alpha},\bm{\beta}, \bm{P}\right)$ is defined for expression brevity.
	\begin{figure*}[hb]
		\centering
		\hrulefill
		\begin{equation}
		\label{Lagrangian}
		\begin{aligned}
			L\left(\bm{\alpha},\bm{\beta}, \bm{P},\bm{\tau},\bm{\rho}\right)&= \sum_{i \in \mathcal{M}}\sum_{j \in \mathcal{M}_{i}}\beta_{i,j}V_{i,j}^{S}+\sum_{i \in \mathcal{M}}\tau_{i}\left(\delta_{0}-\sum_{j \in \mathcal{M}_{i}}\beta_{i, j}\delta_{i, j}\right)+\sum_{i \in \mathcal{M}}\rho_{i}\left(\sum_{j \in \mathcal{M}_{i}}\beta_{i, j}V_{i,j}^{S}-V_{0}\right)\\
			&=\sum_{i \in \mathcal{M}}\sum_{j \in \mathcal{M}_{i}}\beta_{i,j}\left[\left(1+\rho_{i}\right)V_{i,j}^{S}-\tau_{i}\delta_{i, j}\right]+\delta_{0}\sum_{i \in \mathcal{M}}\tau_{i}-V_{0}\sum_{i \in \mathcal{M}}\rho_{i}\\
			&\triangleq \widetilde{L}_{\bm{\tau},\bm{\rho}}\left(\bm{\alpha},\bm{\beta}, \bm{P}\right)+\delta_{0}\sum_{i \in \mathcal{M}}\tau_{i}-V_{0}\sum_{i \in \mathcal{M}}\rho_{i}.
		\end{aligned}
		\end{equation}
	\end{figure*}
	Then, the Lagrange dual problem of $\mathbf{P0}$ becomes
	 \begin{align}
			\mathbf{D0}:\ \min_{\bm{\tau},\bm{\rho}} \quad & D\left(\bm{\tau},\bm{\rho}\right)=g_{\bm{\alpha},\bm{\beta},\bm{P}}\left(\bm{\tau},\bm{\rho}\right)\!+\!\delta_{0}\!\sum_{i \in \mathcal{M}}\tau_{i}\!-\!V_{0}\!\sum_{i \in \mathcal{M}}\rho_{i}~\label{D}\\
			{\rm s.t.} \quad & \tau_{i}\geqslant 0,\ \rho_{i}\geqslant 0,\ \forall i \in \mathcal{M},\tag{\ref{D}a}
	\end{align}
	where
	\begin{equation}
		\label{Dual}
			\begin{aligned}
			g_{\bm{\alpha},\bm{\beta},\bm{P}}\left(\bm{\tau},\bm{\rho}\right) \ &= \ \sup_{\bm{\alpha},\bm{\beta}} \ \widetilde{L}_{\bm{\tau},\bm{\rho}}\left(\bm{\alpha},\bm{\beta}, \bm{P}\right)\\
			{\rm s.t.} \ & \ \text{(\ref{P0}a)}-\text{(\ref{P0}d)}, \text{(\ref{P0}g)}-\text{(\ref{P0}i)}.
			\end{aligned}
	\end{equation}
	Notably, the optimality of the convex problem $\mathbf{D0}$ gives at least the best upper bound of $\mathbf{P0}$, even if $\mathbf{P0}$ is nonconvex, according to the duality property~\cite{boyd2004convex}.
	Hence, our focus now naturally shifts to seeking the optimal solution to $\mathbf{D0}$. 
	
	Given the initial dual variable $\bm{\tau}$ and $\bm{\rho}$, we can solve problem~(\ref{Dual}) in the first place to find the optimal solution of $(\bm{\alpha}, \bm{\beta}, \bm{P})$, the details of which will be presented in the subsequent subsections.
	After that, a subgradient method is employed for updating $\bm{\tau}$ and $\bm{\rho}$ to solve $\mathbf{D0}$ in an iterative fashion.
	Particularly, the partial derivatives with respect to (w.r.t.) $\bm{\tau}$ and $\bm{\rho}$ in $D\left(\bm{\tau},\bm{\rho}\right)$ are set as the subgradient directions, respectively.
	Suppose in a certain iteration, say iteration $t$, each dual variable $\tau_{i}(t)$ ($i \in \mathcal{M}$) is updated by
	\begin{equation}
		\tau_{i}(t+1)=\left[\tau_{i}(t)-\nu_{1}(t)\cdot \left(\delta_{0}-\sum_{j \in \mathcal{M}_{i}}\beta_{i, j}(t)\delta_{i, j}(t)\right)\right]^{+},\label{lagupdate}
	\end{equation}
	and each $\rho_{i}(t)$ is updated by
	\begin{equation}
		\rho_{i}(t+1)=\left[\rho_{i}(t)+\nu_{2}(t)\cdot \left(V_{0}-\sum_{j \in \mathcal{M}_{i}}\beta_{i, j}(t)V_{i,j}^{S}(t)\right)\right]^{+}.\label{lagupdate2}
	\end{equation}
	$\nu_{1}(t)$ and $\nu_{2}(t)$ are the stepsizes w.r.t. the update of $\tau_{i}(t)$ and $\rho_{i}(t)$ in iteration $t$, respectively.
	Generally, the convergence of the subgradient descent method can be ensured with the proper stepsize preset in practice~\cite{boyd2003subgradient}.

	\subsection{Dual Problem Decomposition for Optimality Guarantee}
	As discussed earlier, given $\bm{\tau}$ and $\bm{\rho}$ in each iteration, the optimal $(\bm{\alpha}, \bm{\beta}, \bm{P})$ need to be obtained by solving problem~(\ref{Dual}).
    However, solving such a problem is still challenging due to the inseparability of these variables $\widetilde{L}_{\bm{\tau},\bm{\rho}}\left(\bm{\alpha},\bm{\beta}, \bm{P}\right)$.
    To this end, we propose a two-stage method to reach its optimality with a low computational complexity.
    
    In the first stage, we focus on multiple independent KB caching subproblems, each corresponding to a potential D2D SU pair in the SSCN.
	Specifically, the performances of an SU $i$-SU $j$ single pair (i.e., the sender SU $i$ and the receiver SU $j$) and an SU $j$-SU $i$ pair (i.e., the sender SU $j$ and the receiver SU $i$) should be jointly considered, and for ease of distinction, we refer to the two case as the same~\textit{SU $i,j$ pair}, $\forall \left( i,j\right) \in \mathcal{M}\times \mathcal{M}_{i}, j>i$.
	In other words, for each KBC subproblem, we have $\beta_{i, j}=\beta_{j, i}=1$ in $\widetilde{L}_{\bm{\tau},\bm{\rho}}\left(\bm{\alpha},\bm{\beta}, \bm{P}\right)$ corresponding to a given SU $i,j$ pair, while all other SU pairs are not taken into account.
	Therefore, different KBC subproblems can be tackled independently, and in this way, let
	\begin{equation}
			\omega_{i,j}=\left[\left(1+\rho_{i}\right)V_{i,j}^{S}-\tau_{i}\delta_{i, j}\right]+\left[\left(1+\rho_{j}\right)V_{j,i}^{S}-\tau_{j}\delta_{j, i}\right].~\label{SUpaircost}
	\end{equation}
	Clearly, we have $\omega_{i,j}=\omega_{j,i}$, thus only the case of $j>i$ needs to be considered for each potential SU $i,j$ pair, and constraints (\ref{P0}c), (\ref{P0}d), and (\ref{P0}h) are satisfied at the same time.
	
	Clearly, we have $U=\left(\sum_{i \in \mathcal{M}}\left|\mathcal{M}_{i}\right|\right)/2$ subproblems in total, each of which is denoted as $\mathbf{P1}_{i,j}$, $\forall \left( i,j\right) \in \mathcal{M}\times \mathcal{M}_{i}, j>i$, to seek the optimal KBC sub-policy only for an individual SU $i,j$ pair.
	Note that the optimal KBC solution to problem~(\ref{Dual}) cannot be achieved by simply combining the obtained sub-policies of these $\mathbf{P1}_{i,j}$, but these sub-policies will be used to construct the subsequent DUP and power allocation subproblems to finalize the joint optimal solution of $(\bm{\alpha}, \bm{\beta}, \bm{P})$ for~(\ref{Dual}).
	Given the dual variable $\bm{\tau}$ and $\bm{\rho}$ in each iteration, $\mathbf{P1}_{i,j}$ becomes
	\begin{align}
		\mathbf{P1}_{i,j}:\ \max_{\left\{\alpha_{i}^{k}\right\},\left\{\alpha_{j}^{k}\right\}, P_{i}, P_{j}}\quad &\omega_{i,j}~\label{P1u}\\
		{\rm s.t.} \quad \quad \ & \sum_{k \in \mathcal{K}} \alpha_{i}^{k}\cdot s_{k}\leqslant C_{i},\tag{\ref{P1u}a}\\
		&\sum_{k \in \mathcal{K}} \alpha_{j}^{k}\cdot s_{k}\leqslant C_{j},\tag{\ref{P1u}b}\\
		&\eta_{i}\geqslant\eta_{0},\ \eta_{j}\geqslant\eta_{0},\tag{\ref{P1u}c}\\
		& \alpha_{i}^{k}\in \left\{ 0,1\right\}, \alpha_{j}^{k}\in \left\{ 0,1\right\}, \forall k \in \mathcal{K},\tag{\ref{P1u}d}\\
		& 0\leqslant P_{i}\leqslant P_{max}, \ 0\leqslant P_{j}\leqslant P_{max}.\tag{\ref{P1u}e}
	\end{align}
	If we can obtain the optimal KBC and power allocation sub-policies by solving $\mathbf{P1}_{i,j}$ for SU $i$ (denoted as $\bm{\alpha}^{*}_{i_{(j)}}$ and $P^{*}_{i_{(j)}}$) and SU $j$ (denoted as $\bm{\alpha}^{*}_{j_{(i)}}$ and $P^{*}_{j_{(i)}}$),\footnote{For auxiliary illustration, we use $(\cdot)$ in the subscript to specify the SU pair attribute (relation) for each SU's KBC and power allocation sub-policies obtained from $\mathbf{P1}_{i,j}$.} corresponding to the individual SU $i,j$ pair.
	The following proposition explicitly shows how $\mathbf{P1}_{i,j}$ correlates to the problem~(\ref{Dual}).
	\begin{myPropos}
		Let $\bm{\alpha}^{*}=\left[\bm{\alpha}^{*}_{1}, \bm{\alpha}^{*}_{2}, \cdots,\bm{\alpha}^{*}_{M}\right]^{T}$ be the optimal KBC solution and $\bm{P}^{*}=\left[P_{1}^{*},P_{2}^{*}, \cdots,P_{M}^{*}\right]^{T}$ be the optimal power allocation solution to the problem in~(\ref{Dual}) given dual variables $\bm{\tau}$ and $\bm{\rho}$, where $\bm{\alpha}^{*}_{i}$ and $P_{i}^{*}$ represent the optimal KBC policy and power allocation policy of SU $i$, respectively.
		Then we have $\forall i \in \mathcal{M}$, $\exists j \in \mathcal{M}_{i}$, such that $\bm{\alpha}^{*}_{i_{(j)}}=\bm{\alpha}^{*}_{i}$ and $P^{*}_{i_{(j)}}=P_{i}^{*}$.
	\end{myPropos}
	\begin{IEEEproof}
			Please see Appendix A.
	\end{IEEEproof}
	
	Based on Proposition 1, it is seen that the optimal KBC and power allocation policies of each SU can be obtained by solving a certain $\mathbf{P1}_{i,j}$.
	Moreover, due to the single-association requirement of D2D SemCom pairing, the optimal DUP strategy becomes the only key to finalize the optimal solution to~(\ref{Dual}).
	Hence, we first construct the optimal coefficient matrix for $\bm{\beta}$ in~(\ref{Dual}) to account for all DUP possibilities.
	By calculating the optimal $\omega_{i,j}$ (denoted as $\omega_{i,j}^{*},\forall \left( i,j\right) \in \mathcal{M}\times \mathcal{M}_{i}$) in $\mathbf{P1}_{i,j}$, the optimal coefficient matrix is given by
	\begin{equation}
		\label{optimatrix}
		\bm{\Omega}=\left[
			\begin{matrix}
				+\infty &\omega_{1,2}^{*}&\omega_{1,3}^{*}&\cdots & \omega_{1,M}^{*}\\
				\omega_{2,1}^{*} &+\infty&\omega_{2,3}^{*}&\cdots & \omega_{2,M}^{*} \\
				\omega_{3,1}^{*} &\omega_{3,2}^{*}&+\infty&\cdots & \omega_{3,M}^{*} \\
				\vdots & \vdots & \vdots &\ddots & \vdots\\
				\omega_{M,1}^{*} & \omega_{M,2}^{*} & \omega_{M,3}^{*}& \cdots & +\infty
			\end{matrix}
			\right].
	\end{equation}
	$\bm{\Omega}$ is an $M\times M$ symmetric matrix in which $\omega_{i,j}^{*}=\omega_{j,i}^{*}$, and all elements on its main diagonal are set to $+\infty$ to indicate $i\neq j$.
	In addition, some $\omega_{i,j}^{*}$s in $\bm{\Omega}$ may have a value of $+\infty$ when SU $j$ is not the eligible neighbor of SU $i$, i.e., $j \notin \mathcal{M}_{i}$.
	
	On this basis, we will focus on finding the optimal DUP strategy by constructing a related subproblem in the second stage.
	With the objective $\widetilde{L}_{\bm{\tau},\bm{\rho}}\left(\bm{\alpha},\bm{\beta}, \bm{P}\right)$ and $\bm{\beta}$-related constraints in~(\ref{Dual}), the DUP subproblem is constructed as
	\begin{align}
		\mathbf{P2}:\ \max_{\bm{\beta}}\quad &\frac{1}{2}\sum_{i \in \mathcal{M}}\sum_{j \in \mathcal{M}_{i}}\beta_{i, j}\omega_{i,j}^{*}~\label{P2}\\
		{\rm s.t.} \quad & \text{(\ref{P0}c)},\text{(\ref{P0}d)}, \text{(\ref{P0}h)}.\tag{\ref{P2}a}
	\end{align}
	Given any $\bm{\tau}$ and $\bm{\rho}$, the optimal DUP strategy $\bm{\beta}$ (denoted as $\bm{\beta}^{*}=\left[\beta_{1, j^{*}_{1}},\beta_{2, j^{*}_{2}},\cdots,\beta_{M, j^{*}_{M}}\right]^{T}$) can be straightforwardly finalized by solving $\mathbf{P2}$, where $\beta_{i, j^{*}_{i}}$ ($\forall i \in \mathcal{M}$) means that SU $j^{*}_{i}$ is the optimal D2D SemCom node for SU $i$, i.e., $\beta_{i, j^{*}_{i}}=1$.
	Then, we feed the obtained $\bm{\beta}^{*}$ back to $\bm{\Omega}$ to further determine the optimal KBC and power allocation policies $\bm{\alpha}^{*}$ and $\bm{P}^{*}$ for all SUs.
	In other words, for any $i \in \mathcal {M}$, we have $\bm{\alpha}_{i}^{*}=\bm{\alpha}^{*}_{i_{(j^{*}_{i})}}$ and $P_{i}^{*}=P^{*}_{i_{(j^{*}_{i})}}$, which is established by giving the following proposition.
	\begin{myPropos}
		Given any dual variable $\bm{\tau}$ and $\bm{\rho}$, $\left(\bm{\alpha}^{*},\bm{\beta}^{*},\bm{P}^{*}\right)$ is exactly the optimal solution to problem~(\ref{Dual}).
	\end{myPropos}
	\begin{IEEEproof}
			Please see Appendix B.
	\end{IEEEproof}
	
     Based on Proposition 2, we have proved that the proposed two-stage method is guaranteed to reach the optimality of problem~(\ref{Dual}) in each iteration,.
     Most importantly, the optimization to either $\mathbf{P1}_{i,j}$ or $\mathbf{P2}$ becomes quite tractable owing to the significantly reduced number of variables.
     In the next two subsections, we will elucidate our optimal solutions to $\mathbf{P1}_{i,j}$ and $\mathbf{P2}$, respectively.
	
	\subsection{Near-Optimal KBC and Power Allocation for A D2D Pair}
	Carefully examining $\mathbf{P1}_{i,j}$, it is seen that the two binary variables $\bm{\alpha}_{i}$ and $\bm{\alpha}_{j}$ are the main tricky points during the solving process.
	Suppose that we fix $\bm{\alpha}_{i}$ and $\bm{\alpha}_{j}$ to any feasible solutions in the first place, only constraint (\ref{P1u}e) remains and then we just need to concentrate upon its objective function $\omega_{i,j}$ w.r.t. variable $P_{i}$ and $P_{j}$.
	Note that $\delta_{i,j}$ and $\delta_{j,i}$ in $\omega_{i,j}$ are monotonically increasing functions on $P_{i}$ and $P_{j}$, respectively, which fact is quite obvious by observing \eqref{Qdelay}.
	Moreover, when $\bm{\alpha}_{i}$ and $\bm{\alpha}_{j}$ are fixed to be known, $V_{i,j}^{S}$ can be rephrased to a more precise form, that is
	\begin{equation}
		V_{i,j}^{S} = \left[\theta_{i,j}^{D}\log_{2}\left(1+\frac{P_{i}G_{i,j}^{D}}{\delta^{2}}\right)-\theta_{i}^{E}\log_{2}\left(1+\frac{P_{i}G_{i}^{E}}{\delta^{2}}\right)\right]^{+},\label{sfwa}
	\end{equation}
	where $\theta_{i,j}^{D}$ and $\theta_{i}^{E}$ are constant parameters obtained by substituting the fixed $\bm{\alpha}_{i}$ and $\bm{\alpha}_{j}$ into \eqref{saww} and \eqref{saww2}, respectively.
	From \eqref{sfwa}, we can easily calculate $V_{i,j}^{S}$'s first order derivative w.r.t. $P_{i}$ and obtain its monotonicity between $0$ and $P_{max}$.
	Likewise, the monotonicity of $P_{j}$ in $V_{j,i}^{S}$ can be easily determined at the same time.
	As such, since $V_{i,j}^{S}$ and $V_{j,i}^{S}$ are convex functions while $\delta_{i,j}$ $\delta_{j,i}$ are linear functions, their linear combination $\omega_{i,j}$ must also be convex given any $\bm{\rho}$ and $\bm{\tau}$.
	Therefore, if $\bm{\alpha}_{i}$ and $\bm{\alpha}_{j}$ can be fixed first, we can directly apply efficient linear programming toolboxes like CVXPY~\cite{diamond2016cvxpy} to find the optimum $P_{i}$ and $P_{j}$ for the best $\omega_{i,j}$.
	As such, we propose a heuristic KBC search algorithm by drawing on tabu search~\cite{glover1989tabu} to efficiently determine the near-optimal solution for each $\mathbf{P1}_{i,j}$.
	In detail, the search process is illustrated as follows:
	\begin{itemize}
		\item \textit{Initial KBC Solution Generation:}
		In the beginning, we first determine a feasible solution of $\bm{\alpha}_{i}$ and $\bm{\alpha}_{j}$ (jointly denoted as a $2K$-dimensional vector $\bm{\alpha}_{i,j}^{I}=\left[\bm{\alpha}_{i}^{I},\bm{\alpha}_{j}^{I}\right]$) as the search starting point, where $\bm{\alpha}_{i}^{I}$ and $\bm{\alpha}_{j}^{I}$ are two $K$-dimensional vectors with all elements being $0$ initialized to represent the KBC sub-policies at SU $i$ and SU $j$, respectively.
		Considering constraint (\ref{P1u}c), let $\hat{\mathcal{K}}$ and $\check{\mathcal{K}}$ denote two variable sets to record $\eta_{i}$-relevant information, where $\hat{\mathcal{K}}=\mathcal{K}$ and $\check{\mathcal{K}}=\emptyset$ are initialized.
		Having these, we determine a certain KB $k_{0}$ that leads to the highest sum of KB preferences of SU $i$ and SU $j$ by
		\begin{equation}
			k_{0}=\arg \max_{k\in \hat{\mathcal{K}}}\ \left(p_{i}^{n}+p_{j}^{n}\right),\label{premax}
		\end{equation}
		such as
		\begin{equation}
			\alpha_{i}^{k_{0}}=1 \quad \text{and}\quad \alpha_{j}^{k_{0}}=1.\label{premax2}
		\end{equation}
		Then, let $\hat{\mathcal{K}}=\hat{\mathcal{K}}\backslash n_{0}$ and $\check{\mathcal{K}}=\check{\mathcal{K}}\cup\{n_{0}\}$, and repeat the two procedures in~(\ref{premax}) and~(\ref{premax2}) until both SUs meet constraint~(\ref{P1u}c).
		However, it should note that constraint~(\ref{P1u}a) or (\ref{P1u}b) may be violated during the above process.
		Once that happens, the corresponding KBC indicator of the KB with the maximum size in $\check{\mathcal{K}}$, i.e.,
		\begin{equation}
			k_{1}=\arg \max_{k\in \check{\mathcal{K}}}\ s_{k},
		\end{equation}
		should be reset by
		\begin{equation}
			\alpha_{i}^{k_{1}}=0 \quad \text{and}\quad \alpha_{j}^{k_{1}}=0.\label{Inisolu}
		\end{equation}
		As a result, we can finally generate an initial feasible solution of $\bm{\alpha}_{i,j}^{I}$ to problem $\mathbf{P1}_{i,j}$.
		Meanwhile, as mentioned earlier, the optimal $P_{i}^{I}$ and $P_{j}^{I}$ corresponding to $\bm{\alpha}_{i,j}^{I}$ can be obtained.
	
		\item \textit{Neighboring Space Exploration:}
		Our search algorithm is started from all solutions neighboring to the obtained $\bm{\alpha}_{i,j}^{I}$, and let $\mathcal{H}\left(\cdot\right)$ denote the neighboring solution set of its input solution.
		Next, we need to find the optimal solution within $\mathcal{H}\left(\bm{\alpha}_{i,j}^{I}\right)$ that can yield the minimum value of $\omega_{i,j}$, and then replacing the previous solution $\bm{\alpha}_{i,j}^{I}$ as well as $P_{i}^{I}$ and $P_{j}^{I}$ with the current solution, denoted by $\left(\bm{\alpha}_{i,j}^{C}, P_{i}^{C},P_{j}^{C}\right)$.
		By repeating the above search process in an iteration fashion, we can obtain multiple different current solution in different search iterations.
		Besides, let $\mathcal{I}$ denote an optimal solution list with a given maximum length, and once we have a new better $\bm{\alpha}_{i,j}^{C}$ from searching $\mathcal{H}\left(\bm{\alpha}_{i,j}^{C}\right)$, the solution of $\left(\bm{\alpha}_{i,j}^{C}, P_{i}^{C},P_{j}^{C}\right)$ should be added into $\mathcal{I}$ if it is not already in $\mathcal{I}$.
		The existence of $\mathcal{I}$ is to prevent each search from looping back to previously visited solution spaces, i.e., avoiding trapping into the local optimum.
		Technically, $\mathcal{H}\left(\bm{\alpha}_{i,j}^{C}\right)$ is given as
		\begin{equation}
			\begin{split}
    			\mathcal{H}(\bm{\alpha}_{i,j}^{C}) = \bigl\{ \bm{\alpha}_{i,j}\colon & \|\bm{\alpha}_{i,j}-\bm{\alpha}_{i,j}^{C}\| \leqslant \sigma, \\
    			&\bm{\alpha}_{i,j} \notin \mathcal{I}, \bm{\alpha}_{i,j} \in \psi \bigr\},\label{neighborhood}
			\end{split}
		\end{equation}
		where $\sigma$ is the maximum neighboring length, and $\psi$ is $\mathbf{P1}_{i,j}$'s entire feasible solution space.
		\item \textit{Optimal Solution Update and Termination Check:}
		Let $\left(\bm{\alpha}_{i,j}^{*},P_{i}^{*},P_{j}^{*}\right)$ denote a variable vector to record the best solution obtained so far.
		Specifically, if we find the current $\left(\bm{\alpha}_{i,j}^{C}, P_{i}^{C},P_{j}^{C}\right)$ that is better than $\left(\bm{\alpha}_{i,j}^{*},P_{i}^{*},P_{j}^{*}\right)$ in any iteration, this should not be added into $\mathcal{I}$ but let
		\begin{equation}
			\bm{\alpha}_{i,j}^{*}=\bm{\alpha}_{i,j}^{C}, \quad P_{i}^{*}=P_{i}^{C},\quad \text{and}\quad P_{j}^{*}=P_{j}^{C}.
		\end{equation}
		Moreover, a search termination criterion is checked by either reaching a maximum number of iterations or a performance growth threshold in terms of $\omega_{i,j}$.
	\end{itemize}
	
	\subsection{Optimal DUP Scheme}
	After solving each $\mathbf{P1}_{i,j}$, the optimal coefficient matrix $\bm{\Omega}$ as well as $\mathbf{P2}$ can be determined.
	Clearly, its objective function and constraints~(\ref{P0}c) and (\ref{P0}d) are all linear, the only challenge to solve $\mathbf{P2}$ is the $0$-$1$ constraint in~(\ref{P0}h).
	Herein, we relax $\bm{\beta}$ into the continuous variable, denoted as $\bm{\beta}^{R}$, between $0$ and $1$, which makes $\mathbf{P2}$ a linear programming problem that is able to be efficiently tackled by CVXPY.
	Afterward, we need to recover $\bm{\beta}^{R}$ to its binary property, and for this, we propose a heuristic DUP strategy as follows. First, we determine one single SU $i'$-SU $j'$ pair by
	\begin{equation}
	\label{VSPsolution}
		\beta_{i^{'}, j^{'}}^{*}=\beta_{j^{'}, i^{'}}^{*}=1
	\end{equation}
	if
	\begin{equation}
		\label{VSPsolution1}
		\left(i^{'},j^{'}\right)=\arg \max_{i\in \mathcal{M},j\in \mathcal{M}_{i},j>i}\beta_{i, j}^{R}.
	\end{equation}
	For the remaining $\beta_{i, j}^{*}$ w.r.t. SU $i^{'}$ and SU $j^{'}$, we have
	\begin{equation}
		\label{VSPsolution2}
			\left\{
			\begin{aligned}
			\beta_{i^{'}, j}^{*}=\beta_{j, i^{'}}^{*}=0, \quad & \forall j \in \mathcal{M}_{i^{'}},j \neq j^{'}\\
			\beta_{j^{'}, i}^{*}=\beta_{i, j^{'}}^{*}=0, \quad & \forall i \in \mathcal{M}_{j^{'}},i \neq i^{'}
			\end{aligned}.
		\right.
	\end{equation}
	Then we let $\mathcal{M}=\mathcal{M}\backslash \{i,j\}$, and repeat the above processes until finalizing the optimal DUP solutions for all SU pairs.
	It is observed that the number of variables is only $\left(\sum_{i \in \mathcal{M}}\left|\mathcal{M}_{i}\right|\right)/2$ in solving $\mathbf{P2}$, which is a fairly acceptable problem scale in practice, and the performance compromise is believed to be small.
	
	\subsection{Algorithm Analysis}
	To better demonstrate the full picture of the proposed solution, we summarize the relevant technical points and enclose them in the following Algorithm~\ref{Algo1}.
	\begin{breakablealgorithm}
				\caption{The Proposed Resource Management for SSCN}
				\label{Algo1}
				\begin{algorithmic}[1]
					\REQUIRE \textit{The system parameters $M$, $K$, $W$, $L$, $\zeta$, $P_{max}$, $\gamma_{0}$, $\eta_{0}$, $\delta_{0}$, $V_{0}$, and the parameters of the eavesdropping center and all SUs $G_{i,j}^{D}$, $G_{i}^{E}$, $s_{k}$, $C_{i}$, $r_{i}^{k}$, $r_{E}^{k}$, $\xi_{i}$, $\xi_{E}$, $\mu_{j}^{k}$}
					\ENSURE \textit{The optimal KBC strategy $\bm{\alpha}^{*}$, optimal DUP strategy $\bm{\beta}^{*}$, and optimal power allocation strategy $\bm{P}^{*}$}
					\STATE \textit{Initialize dual problem iteration index $t\leftarrow 1$, $\tau_{i}(1)$, $\rho_{i}(1)$, $\nu_{1}(1)$, and $\nu_{2}(1)$ to proper positive values for $\mathbf{D0}$}
					\STATE \textit{Set the maximum number of iterations $Q$}
					\WHILE{$t\leqslant Q$}
						\FOR{\textit{$i \leftarrow1$ to $M$}}
							\FOR{\textit{$j \leftarrow1$ to $M$}}
							\IF{\textit{$j \in \mathcal{M}_{i}$ and $j>i$}}
									\STATE \textit{Find $\left(\bm{\alpha}_{i,j}^{I}, P_{i}^{I},P_{j}^{I}\right)$ in the context of (\ref{premax})-(\ref{Inisolu})}
									\STATE \textit{Initialize KBC search iteration index as $t'\leftarrow1$, $\mathcal{I}(1)\leftarrow\emptyset$, $\bm{\alpha}_{i,j}^{C}(1)\leftarrow\bm{\alpha}_{i,j}^{*}\leftarrow\bm{\alpha}_{i,j}^{I}$, $P_{i}^{C}(1)\leftarrow P_{i}^{*}\leftarrow P_{i}^{I}$, and $P_{j}^{C}(1)\leftarrow P_{j}^{*}\leftarrow P_{j}^{I}$}
									\STATE \textit{Set a suitable maximum neighboring length $\sigma$ and maximum number of iterations $Q'$ for $\mathbf{P1}_{i,j}$}
									\WHILE{$t'\leqslant Q'$}
										\STATE \textit{Determine $\mathcal{H}\left(\bm{\alpha}_{i,j}^{C}\left(t'\right)\right)$ by (\ref{neighborhood})}
										\STATE \textit{Find the best KBC solution in $\mathcal{H}\left(\bm{\alpha}_{i,j}^{C}\left(t'\right)\right)$ and then obtain the corresponding optimal power allocation solution by CVXPY}
										\STATE \textit{Assign them to $\bm{\alpha}_{i,j}^{C}(t'+1)$, $P_{i}^{C}(t'+1)$, and $P_{j}^{C}(t'+1)$, respectively}
										\IF{\textit{$\left(\bm{\alpha}_{i,j}^{C}(t'+1), P_{i}^{C}(t'+1),P_{j}^{C}(t'+1)\right)$ is better than $\left(\bm{\alpha}_{i,j}^{*},P_{i}^{*},P_{j}^{*}\right)$}}
											\STATE \textit{Update $\left(\bm{\alpha}_{i,j}^{*},P_{i}^{*},P_{j}^{*}\right)\leftarrow\left(\bm{\alpha}_{i,j}^{C}(t'+1), P_{i}^{C}(t'+1),P_{j}^{C}(t'+1)\right)$}
											\STATE \textit{Keep $\mathcal{I}\left(t'+1\right)\leftarrow\mathcal{I}\left(t'\right)$}
										\ELSE
											\STATE \textit{Update $\mathcal{I}\left(t'+1\right)\leftarrow\mathcal{I}\left(t'\right)\cup\left\{\left(\bm{\alpha}_{i,j}^{C}(t'+1), P_{i}^{C}(t'+1),P_{j}^{C}(t'+1)\right)\right\}$}
										\ENDIF
										\STATE \textit{Update $t'\leftarrow t'+1$}
									\ENDWHILE
									\STATE\textit{Calculate $\omega_{i,j}^{*}$ by substituting $\left(\bm{\alpha}_{i,j}^{*},P_{i}^{*},P_{j}^{*}\right)$ into~(\ref{SUpaircost})}
									\STATE \textit{$\omega_{j,i}^{*}\leftarrow\omega_{i,j}^{*}$}
							\ENDIF
							\ENDFOR
						\ENDFOR
						\STATE \textit{Generate $\bm{\Omega}\left(t\right)$ according to~(\ref{optimatrix})}
						\STATE \textit{Solve $\mathbf{P2}$ by CVXPY and obtain $\bm{\beta}^{R}\left(t\right)$}
						\STATE \textit{Finalize $\bm{\beta}^{*}\left(t\right)$ by~(\ref{VSPsolution})-(\ref{VSPsolution2})}
						\STATE \textit{Finalize $\bm{\alpha}^{*}\left(t\right)$ and $\bm{P}^{*}(t)$ by feeding $\bm{\beta}^{*}\left(t\right)$ back to $\bm{\Omega}\left(t\right)$}
						\STATE \textit{Update $\tau_{i}(t+1)$ and $\rho_{i}(t+1)$ by~(\ref{lagupdate}) and \eqref{lagupdate2}, respectively}
						\STATE \textit{Update $\nu_{1}(t+1)$ and $\nu_{2}(t+1)$ under a given rule}
						\STATE $t\leftarrow t+1$
					\ENDWHILE
					\RETURN $\left(\bm{\alpha}^{*},\bm{\beta}^{*},\bm{P}^{*}\right) \leftarrow \left(\bm{\alpha}^{*}(Q),\bm{\beta}^{*}(Q),\bm{P}^{*}(Q)\right)$
			\end{algorithmic}
		\end{breakablealgorithm}
	
	Regarding the computational complexity of Algorithm~\ref{Algo1}, it is first observed that for each $\mathbf{P1}_{i,j}$, we need to obtain all feasible KBC solutions within $\mathcal{H}\left(\bm{\alpha}_{i,j}^{C}\right)$ in any of its iterations.
	Note that $\sigma$ is a very small parameter compared with $N$ in~(\ref{neighborhood}), the complexity of determining the neighborhood solution space is approximately $\mathcal{O}\left(\binom{2K}{\sigma}\right)=\mathcal{O}\left(K^{\sigma}\right)$.
	With the maximum number of search iteration $Q'$, solving each $\mathbf{P1}_{i,j}$ needs complexity $\mathcal{O}\left(Q'K^{\sigma}\right)$.
	Besides, the CVXPY toolbox employed for the relaxed $\mathbf{P2}$ requires $\mathcal{O}\left(M^{4}\right)$ complexity~\cite{lee2019solving} to solve a group of $\left(\sum_{i \in \mathcal{M}}\left|\mathcal{M}_{i}\right|\right)$ DUP variables.
	Since there is a total of $\left(\left(\sum_{i \in \mathcal{M}}\left|\mathcal{M}_{i}\right|\right)/2\right)$ subproblems $\mathbf{P1}_{i,j}$ and one subproblem $\mathbf{P2}$ need to be tackled in each iteration of $\mathbf{D0}$, its corresponding complexity is $\mathcal{O}\left(M^{2}Q'K^{\sigma}+M^{4}\right)$.
	Further combining the maximum number of dual problem iteration $Q$, the proposed Algorithm~\ref{Algo1} has a polynomial-time overall complexity of $\mathcal{O}\left(QM^{2}\left(Q'K^{\sigma}+M^{2}\right)\right)$.
	
	
	\section{Numerical Results and Discussions}
	In this section, numerical evaluations are conducted to demonstrate the performance of our proposed resource allocation solution in the SSCN, where we employ Python 3.7-based PyCharm as the simulator platform and implement it in a workstation PC featuring the AMD Ryzen-9-7900X processor with 12 CPU cores and 128 GB RAM.
	In the basic system setup, we first model a single-cell circular area with a radius of $300$ meters, in which multiple SUs and one eavesdropping center are randomly dropped, while multiple KBs are preset to provide SUs with a variety of distinct SemCom services.
	Correspondingly, we set a uniform KB storage capacity for all SUs, and each of them has a randomly generated storage size.
    For brevity, other relevant simulation parameters not mentioned in the context along with their values are summarized in Table~\ref{SimuPara}.
    \begin{table}[t]
		\centering
		\caption{Simulation Parameters}
		\label{SimuPara}
		\setlength{\tabcolsep}{3pt}
		\renewcommand\arraystretch{1.5}
		\begin{tabular}{|m{5cm}<{\raggedright}|m{3.3cm}<{\raggedright}|}\hline
			\textbf{Parameters} & \textbf{Values} \\ \hline
			Subchannel bandwidth ($W$) & $0.1$ MHz\\ \hline
			Number of SUs ($M$) & $100$ \\ \hline
			Number of KBs ($K$) & $12$ \\ \hline
			Size of KB $k$ ($s_{k}$) & $1\sim 5$ units (randomly)~\cite{xia2023xurllc}\\ \hline
			KB storage capacity of SUs ($C_{i}$) & $24$ units \\ \hline
			Skewness of the Zipf distribution w.r.t. each SU's KB preference ($\xi_{i}$) & $1.2$ \\ \hline
			Average interpretation time of KB $n$-based semantic packets ($1/\mu_{i}^{n}$) & $5\times 10^{-3}\sim 1\times 10^{-2}$ s/packet (randomly)~\cite{xia2023xurllc} \\ \hline
			Maximum transmit power of each SU ($P_{max}$) & $21$ dBm~\cite{pawar2021joint}  \\ \hline
			Noise power ($\zeta^{2}$) & $-111.45$ dBm~\cite{boostanimehr2014unified} \\ \hline
			Path loss model & $34+40\log\left(d\ \text{[m]}\right)$ \cite{ye2018dynamic} \\ \hline
			Average number of bits required for encoding one semantic triplet ($L$) & $800$ bits~\cite{xia2024wireless}\\ \hline
			Minimum semantic knowledge satisfaction threshold ($\eta_{0}$) & $0.5$~\cite{xia2023knowledge} \\ \hline
			Maximum queuing delay threshold ($\delta_{0}$) & $5$ ms~\cite{carlucci2018controlling} \\ \hline
			Minimum SST threshold ($V_{0}$) & $50$~\cite{xia2025joint} \\ \hline
		\end{tabular}
	\end{table}	

	Besides, the preference rankings for all KBs (i.e., $r_{i}^{n}$) at the eavesdropping center and each SU are generated independently and randomly, where their Zipf distributions are assumed to have the same skewness $1.2$.
	The average interpretation time for semantic packets based on each KB $n$ is considered to be the same for all SUs.
	Here, we randomly generate $1/\mu_{i}^{n}$ in a range of $5\times 10^{-3}\sim 1\times 10^{-2}$ s/packet w.r.t. each KB $n$, as the average interpretation time of different KBs-based packets is different from each other.
	Further, the minimum knowledge preference satisfaction threshold $\eta_{0}$, the maximum queuing delay threshold $\delta_{0}$, and the maximum SST threshold $V_{0}$ are prescribed as $0.5$, $5$ ms, and $50$, respectively.
	Notably, all the above parameter values are set by default unless otherwise specified, and all subsequent numerical results are obtained by averaging over a sufficiently large number of trials.
	
	For comparison purposes, here we employ two different benchmark schemes of SemCom-empowered service provisioning herein~\cite{kim2005random,guo2019resource,xia2023xurllc}: 1) Random Power allocation with Distance-first pairing (RPD) strategy, which assumes each SU has a random assigned transmit power while choosing its nearest SU for DUP; 2) Maximum Power allocation with Knowledge-first pairing (MPK), in which each SU is assigned with a transmit power of $P_{max}$ while selecting its neighboring SU with the highest KB matching degree for DUP.
	In addition, a personal preference-first KBC policy is considered for RPD and MPK, which allows each SU to cache KBs with the highest preferences until $\eta_{0}$ is satisfied, and then randomly cache the remaining KBs until reaching maximum capacity.
	
	\begin{figure}[t]
		\centering
		\includegraphics[width=0.45\textwidth]{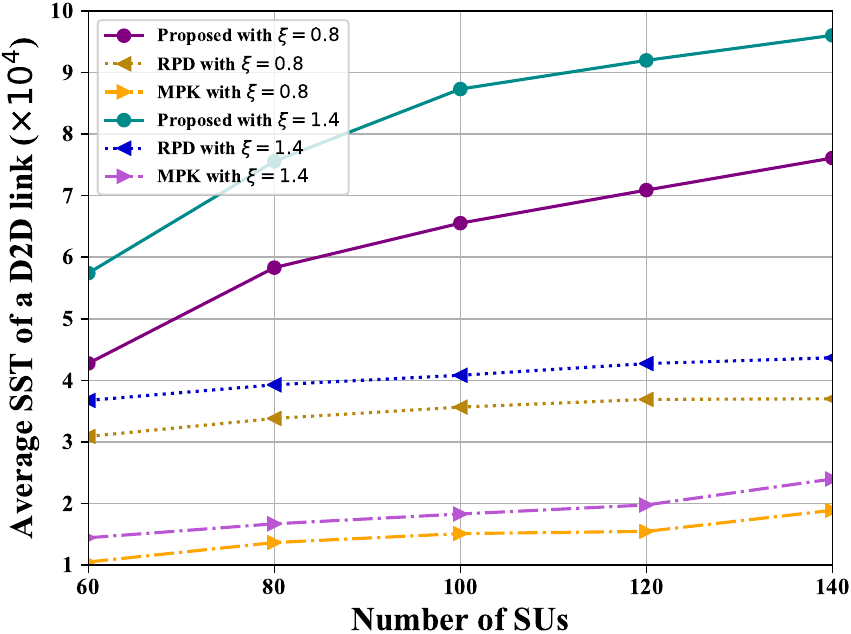} 
		\caption{Average SST vs. different numbers of SUs.}
		\label{sstsu}
    \end{figure}
\begin{figure}[t]
		\centering
		\includegraphics[width=0.45\textwidth]{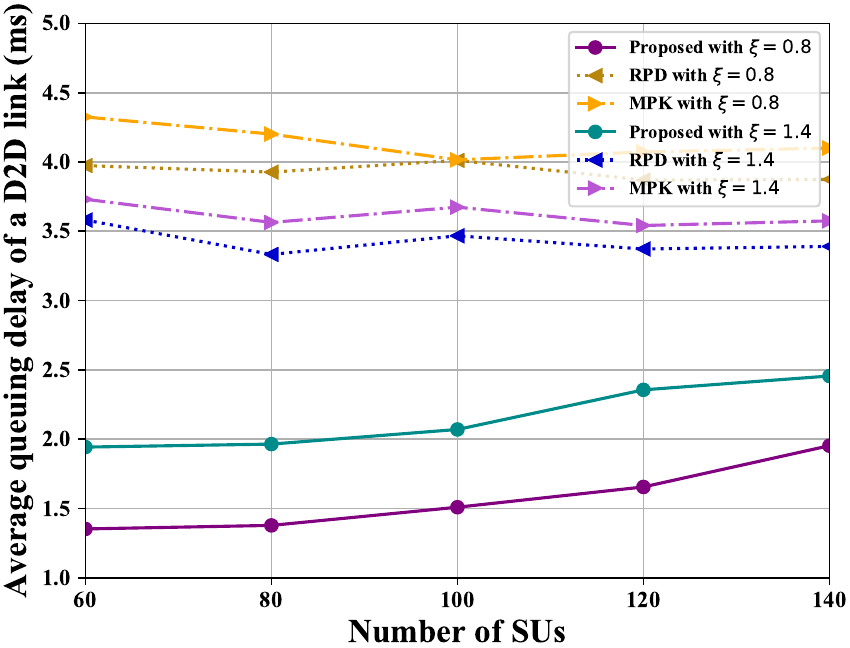} 
		\caption{Average queuing delay vs. different numbers of SUs.}
		\label{delaysu}
    \end{figure}

    We first verify the average SST performance and the average queuing latency performance obtained by each D2D SemCom link under varying numbers of SUs in Fig.~\ref{sstsu} and Fig.~\ref{delaysu}, respectively, in which two different skewness values of the Zipf distribution of $\xi = 0.8$ and $\xi = 1.4$ are taken into account.
    In comparison with the two benchmark schemes, our solution always renders an obvious SST performance gain at any number of SUs, while reaching a positive SST performance representing that the semantic information security has been adequately guaranteed in the SSCN.
    For example, in Fig.~\ref{sstsu}, our solution reaches a high SST performance of around $7.7\times 10^4$ at $140$ SUs with $\xi = 0.8$, which is much higher than around $3.8 \times 10^4$ SST of the RPD and around $1.9 \times 10^4$ SST of the MPK.
    In addition, in Fig.~\ref{delaysu}, a lower queuing latency is seen by our solution, which is less than half of that obtained by the comparison schemes.
    Note that as the increase of the number of SUs, both the SST performance and queuing latency show a grow trend.
    The former is because that the more SUs mean the more neighbors for each SU, and thus each SU should have a higher chance to find a better SemCom counterpart for D2D pairing and to achieve a better SST performance.
     However, such an increase will be eventually stabilized since the bandwidth budget is already fixed.
     Likewise, for the latter queuing delay trend, when each SU has a better semantic value rate, the semantic data packet arrival rate becomes correspondingly higher, thereby leading to a higher queuing delay.
     Furthermore, it is observed that the higher skewness of the Zipf distribution is with the better SST and a worse queuing delay.
     The rationale behind the former phenomenon is that the higher skewness makes each SU easier meet the semantic satisfaction as it only needs to cache the KBs with high probabilities, and the semantic triplets generated based on these KBs precisely have the high semantic value.
     Meanwhile, such a situation happens to cause that each SU is more difficult to find a high knowledge-matching D2D pair due to the higher deviation, hence the queuing delay becomes higher as well compared with the low skewness.

	\begin{figure}[t]
		\centering
		\includegraphics[width=0.45\textwidth]{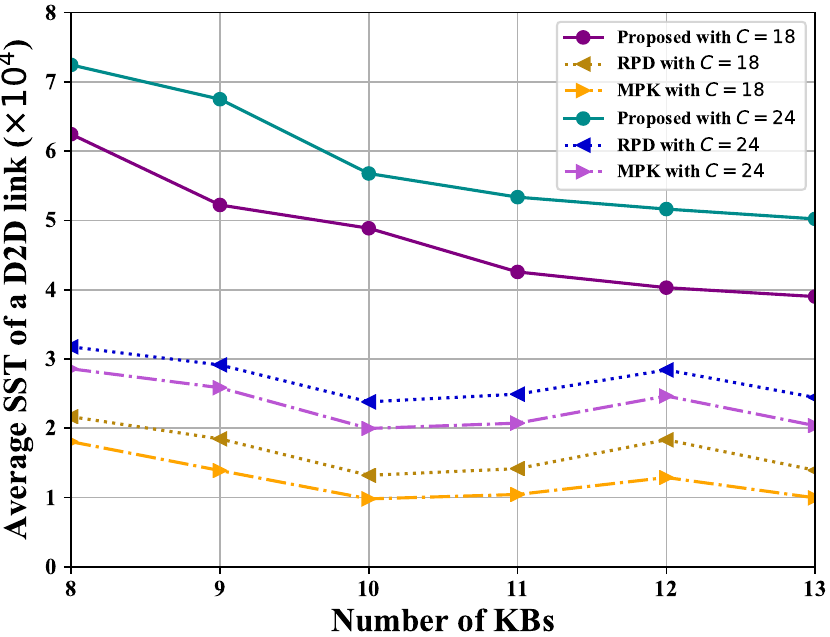} 
		\caption{Average SST vs. different numbers of KBs.}
		\label{sstkb}
    \end{figure}
\begin{figure}[t]
		\centering
		\includegraphics[width=0.45\textwidth]{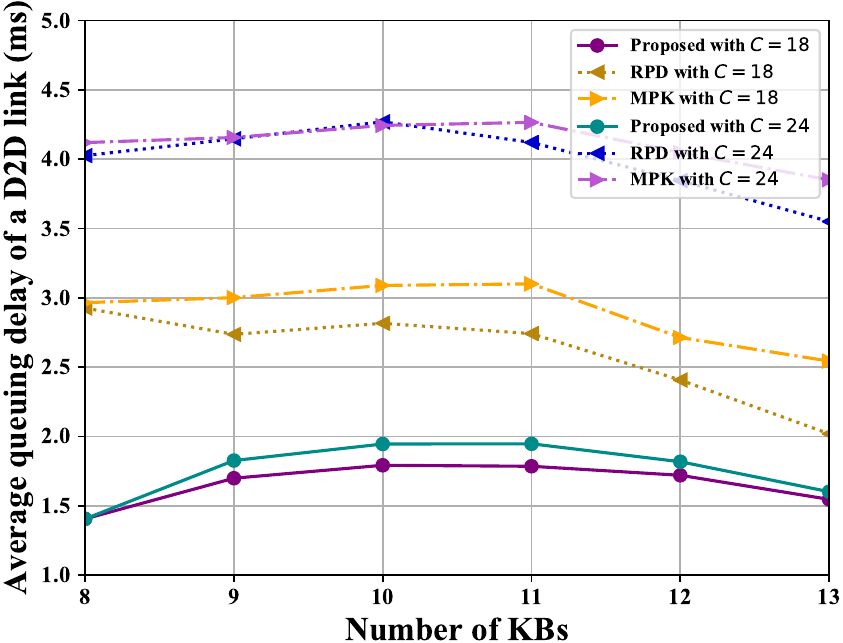} 
		\caption{Average queuing delay vs. different numbers of KBs.}
		\label{delaykb}
    \end{figure}

	Fig.~\ref{sstkb} and~\ref{delaykb} compare the average SST and queuing delay performance of each D2D SemCom link with the benchmarks under varying numbers of KBs, respectively, with two different KB storage capacities of $C=18$ and $C=24$.
	In both figures, it is seen that our proposed solution far superior to either the RPD or the MPK at each point.
	For instance, when the number of KBs is set to $10$ and $C=24$, an average SST performance of $5.8 \times 10^4$ is observed by the proposed solution in Fig.~\ref{sstkb}, which is around $1.9$ times higher than that of the MPK and around $1.4$ times higher than that of the RPD.
	Specially, in Fig.~\ref{sstkb}, our solution shows a downtrend of SST with the increase of the number of KBs.
	This is because that the more the number of KBs, the less the discrepancy of KB preference between SUs and the eavesdropping center, resulting in the higher probability for the eavesdropping center having the same KBs as SUs, which inevitably causes a lower semantic secrecy information rate.
	Besides, a larger KB capacity $C$ has a better SST, which can be interpreted as the more KBs are cached at each SU, the more semantic value is transmitted.

    In Fig.~\ref{delaykb}, an increase trend of the average queuing delay performance is first seen by the proposed solution with the number of KBs, and then decline slowly after $11$ KBs.
    This is attribute to the fact that at the very beginning, the more KBs indicates a higher effective packet arrival rate, which can lead to a grow trend of queuing delay.
    However, such performance saturation will get better as a higher probability for two SUs caching the same KBs with high interpretation rates, which dominates the overall queuing delay performance change.
    Moreover, there is only a small queuing delay gap between the cases with different KB storage capacities.
    This is because that our solution always gives high caching priority to the KBs with lower interpretation times, and once the semantic knowledge satisfaction threshold or the queuing delay threshold is met, the remaining capacity will not be in use.
    
    \begin{figure}[t]
		\centering
		\includegraphics[width=0.45\textwidth]{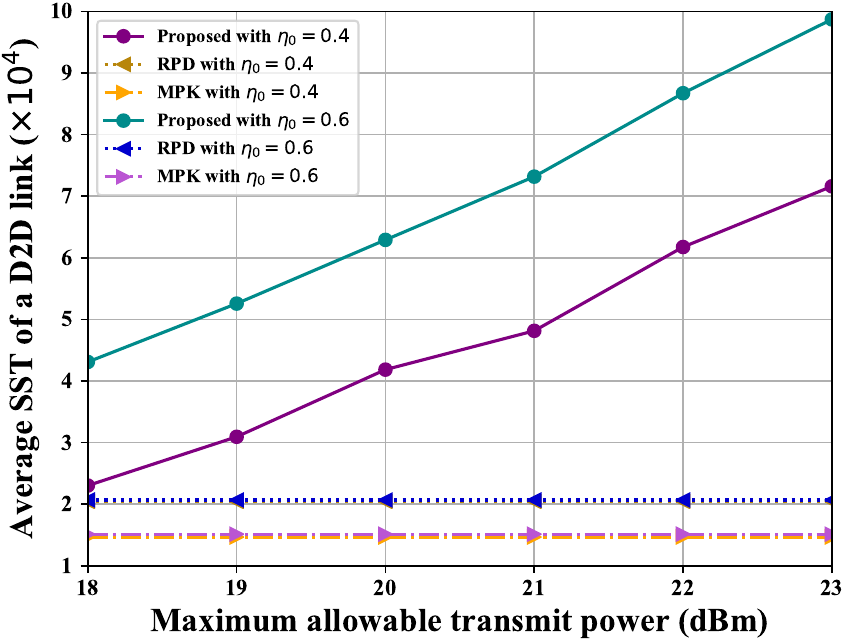} 
		\caption{Average queuing delay vs. different maximum allowable transmit powers.}
		\label{SSTpmax}
    \end{figure}
\begin{figure}[t]
		\centering
		\includegraphics[width=0.45\textwidth]{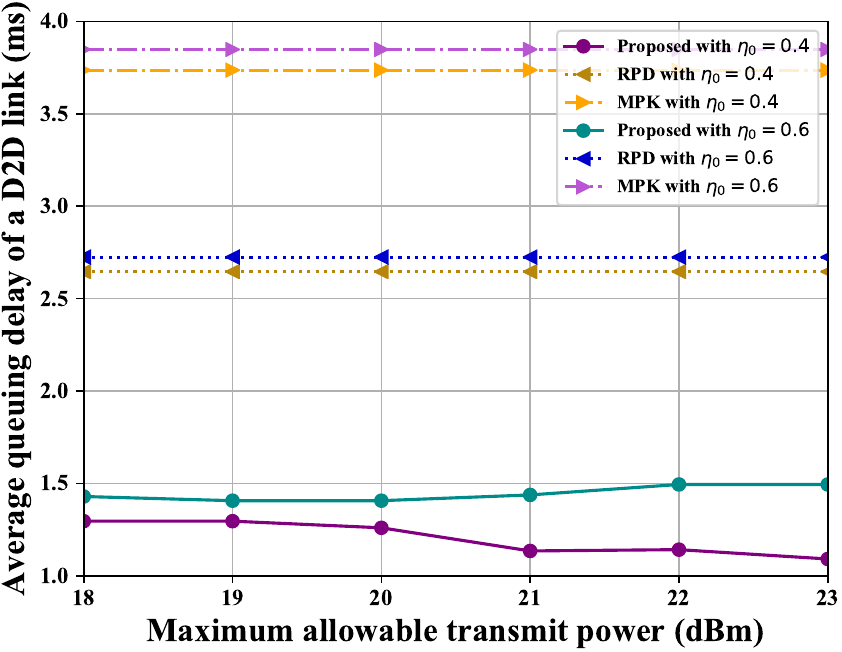} 
		\caption{Average queuing delay vs. different maximum allowable transmit powers.}
		\label{delaypmax}
    \end{figure}

    In addition, Fig.~\ref{SSTpmax} and Fig.~\ref{delaypmax} validate the average queuing delay and SST performance under varying maximum allowable transmit powers, respectively, under two semantic knowledge satisfaction thresholds of $\eta_{0}=0.4$ and $\eta_{0}=0.6$.
    Consistent with the previous results, our solution still outperforms the two benchmarks with a significant performance gain.
    Specifically, at the point of $P_{max}=21$ dBm and $\eta_{0}=0.6$, the proposed solution reaches an average SST performance of around $73000$ in Fig.~\ref{SSTpmax}, which is $53000$ higher than that of RPD, and obtains an average queuing delay of around $1.4$ ms in Fig.~\ref{delaypmax}, which is $2.3$ ms lower than that of MPK.
    Apart from this, note first that in Fig.~\ref{SSTpmax}, the SST performance increases as $P_{max}$.
    This is because that based on our efficient DUP solution, most of the SUs can always find a best SU for D2D SemCom pairing to guarantee a positive SST, in which case each SU is able to be assigned with a highest transmit power to improve the semantic value transmission rate as much as possible.
    Also, the higher $\eta_{0}$ leads to the better SST, which trend is quite obvious due to the looser KBC constraints.
    
    Finally, as in Fig.~\ref{delaypmax}, it is observed that the average queuing latency performance obtained by all resource allocation schemes shows a similar stable trend as the growth of $P_{max}$.
    This phenomenon can be understood by the fact that we just set a queuing latency limitation in the optimization problem, and our main focus is on the maximization of SST rather than minimizing the queuing delay.
    Hence, the corresponding solution here is just to satisfy the delay constraint, resulting in the stable trend.
    Nevertheless, our proposed solution still has a better delay performance compared with the benchmarks, as aforementioned.
    In the meantime, it is noticed that the lower the $\eta_{0}$, the lower the average queuing delay.
    This is because that the lower semantic knowledge satisfaction threshold clearly makes each SU cache less KBs, thus the knowledge overlapping rate between the two paring SUs is more likely to be smaller.
    Consequently, the effective semantic packet arrival rate at a lower $\eta_{0}$ should be smaller than that at a higher $\eta_{0}$, rendering a better queuing latency performance.

	\section{Conclusions}
	This paper has explored the semantic information security-aware resource allocation in the SSCN.
	We have first identified two key problems of KBC and DUP with the joint consideration of power allocation.
	Then, the knowledge matching based queuing delay has been derived and a novel performance metric of SST has been developed.
	On this basis, we have formulated a joint SST maximization problem, followed by a corresponding solution proposed.
	Our solution first leveraged primal-dual transformation and decomposed the dual problem into multiple subproblems, which can be separately solved by our devised two-stage method.
	Compared with the two benchmarks, simulation results have shown superiorities in SST, average queuing delay, and semantic knowledge satisfaction.
	
	We hope this paper can provide some valuable insights for follow-up research on secure SemCom.
	While this work establishes fundamental principles for secure resource allocation in D2D SemCom scenarios, real-world deployment in 5G/6G networks would require addressing additional challenges.
    For instance, in cellular-connected D2D scenarios, our KBC-DUP strategies may be capable of being implemented at edge servers to enable semantic-aware user pairing.
    However, practical constraints such as dynamic KBC across mobile users, semantic protocol standardization gaps, and latency overheads for SST calculation in multi-hop network environments need dedicated investigation, which will be systematically explored through testbed validation in our subsequent work.
	
	\begin{appendices}
		\section{Proof of Proposition 1}
		Given the optimal KBC solution $\bm{\alpha}^{*}$ and optimal power allocation solution $\bm{P}^{*}$, let $\bm{\beta}^{*}=\left[\beta_{1, j^{*}_{1}},\beta_{2, j^{*}_{2}},\cdots,\beta_{M, j^{*}_{M}}\right]^{T}$ be the corresponding optimal DUP solution to the problem in~(\ref{Dual}) under the same dual variable $\bm{\tau}$ and $\bm{\rho}$, where $\beta_{i, j^{*}_{i}}$ ($\forall i \in \mathcal{M}$) indicates that SU $j^{*}_{i}$ is the optimal D2D SemCom counterpart for SU $i$, i.e., $\beta_{i,j^{*}_{i}}=1$.
		
		According to $\omega_{i,j}$ defined in~(\ref{SUpaircost}), the objective function $\widetilde{L}_{\bm{\tau},\bm{\rho}}\left(\bm{\alpha},\bm{\beta}, \bm{P}\right)$ in~(\ref{Dual}) can be rewritten as
		\begin{equation}
		\label{objective1}
			\begin{aligned}
				\widetilde{L}_{\bm{\tau},\bm{\rho}}\left(\bm{\alpha},\bm{\beta}, \bm{P}\right)=&\frac{1}{2}\sum_{i \in \mathcal{M}}\sum_{j \in \mathcal{M}_{i}}\beta_{i, j}\omega_{i,j}=\sum_{i \in \mathcal{M}}\sum_{j \in \mathcal{M}_{i},j>i}\beta_{i,j}\omega_{i,j},
			\end{aligned}
		\end{equation}
		then we substitute $\bm{\beta}^{*}$ into~(\ref{objective1}) and yield
		\begin{equation}
		\label{objective2}
			\begin{aligned}
					\widetilde{L}_{\bm{\tau},\bm{\rho}}(\bm{\alpha},\bm{\beta}^{*},\bm{P})=&\frac{1}{2}\sum_{i \in \mathcal{M}}\omega_{i,j^{*}_{i}}=\sum_{i \in \mathcal{M},j^{*}_{i}>i}\omega_{i,j^{*}_{i}},
			\end{aligned}
		\end{equation}
		where $\omega_{i,j^{*}_{i}}$ is the term only related to SU $i,j^{*}_{i}$ pair.
		
		Clearly, if $\bm{\alpha}^{*}$ and $\bm{P}^{*}$ are further substituted into~(\ref{objective2}), we can directly reach the optimality of the problem~(\ref{Dual}).
		Since different SU $i,j^{*}_{i}$ pairs should be independent of each other in practical systems, it means that different terms of $\omega_{i,j^{*}_{i}}$ are independent of each other as well in $\widetilde{L}_{\bm{\tau},\bm{\rho}}(\bm{\alpha},\bm{\beta}^{*},\bm{P})$.
		Therefore, we are able to conclude that reaching the optimality of $\widetilde{L}_{\bm{\tau},\bm{\rho}}(\bm{\alpha},\bm{\beta}^{*},\bm{P})$ is equivalent to reaching the optimality of each $\omega_{i,j^{*}_{i}}$, where the optimality can be reached when $\bm{\alpha}=\bm{\alpha}^{*}$ and $\bm{P}=\bm{P}^{*}$ are met at the same time.
		As such, $\bm{\alpha}^{*}_{i}$ and $P_{i}^{*}$ must be the optimal KBC and power allocation solutions in terms of the problem to maximize $\omega_{i,j^{*}_{i}}$.
		Further noticing that $\omega_{i,j}$ is the objective of $\mathbf{P1}_{i,j},\forall \left( i,j\right) \in \mathcal{M}\times \mathcal{M}_{i}, j>i$, where $\bm{\alpha}^{*}_{i_{(j)}}$ and $P^{*}_{i_{(j)}}$ are the corresponding optimal solutions, hence the equality $\bm{\alpha}^{*}_{i_{(j)}}=\bm{\alpha}^{*}_{i}$ and $P^{*}_{i_{(j)}}=P_{i}^{*}$ hold when $j=j^{*}_{i}$. This completes the proof.
			
		\section{Proof of Proposition 2}
		If $\left(\bm{\alpha}^{*},\bm{\beta}^{*},\bm{P}^{*}\right)$ is not the optimal solution to~(\ref{Dual}), there must be an optimal solution, denoted as $\bar{\bm{\alpha}}=\left[\bar{\bm{\alpha}}_{1}, \bar{\bm{\alpha}}_{2}, \cdots,\bar{\bm{\alpha}}_{M}\right]^{T}$, $\bar{\bm{\beta}}=\left[\beta_{1, \bar{j}_{1}},\beta_{2, \bar{j}_{2}},\cdots,\beta_{M, \bar{j}_{M}}\right]^{T}$, and $\bar{\bm{P}}=\left[\bar{P}_{1},\bar{P}_{2},\cdots, \bar{P}_{M}\right]^{T}$, such that 
		\begin{equation}
			\widetilde{L}_{\bm{\tau},\bm{\rho}}(\bar{\bm{\alpha}},\bar{\bm{\beta}},\bar{\bm{P}})>\widetilde{L}_{\bm{\tau},\bm{\rho}}(\bm{\alpha}^{*},\bm{\beta}^{*},\bm{P}^{*}).\label{B1}
		\end{equation}
		
		On one hand, as $\bm{\beta}^{*}$ is the optimal solution to $\mathbf{P2}$, for $\bar{\bm{\beta}}\neq \bm{\beta}^{*}$, we have $\widetilde{L}_{\bm{\tau},\bm{\rho}}(\bm{\alpha}^{*},\bm{\beta}^{*},\bm{P}^{*})>\widetilde{L}_{\bm{\tau},\bm{\rho}}(\bm{\alpha}^{*},\bar{\bm{\beta}},\bm{P}^{*})$.
		On the other hand, for the DUP case of $\bm{\beta}=\bar{\bm{\beta}}$, the optimal KBC and power allocation policies have been determined by solving each $\mathbf{P1}_{i,\bar{j}_{i}}$, i.e., $\left(\bm{\alpha}^{*},\bm{P}^{*}\right)$. In other words, we have
		\begin{equation}
			\widetilde{L}_{\bm{\tau},\bm{\rho}}(\bar{\bm{\alpha}},\bar{\bm{\beta}},\bar{\bm{P}})<\widetilde{L}_{\bm{\tau},\bm{\rho}}(\bm{\alpha}^{*},\bar{\bm{\beta}},\bm{P}^{*})<\widetilde{L}_{\bm{\tau},\bm{\rho}}(\bm{\alpha}^{*},\bm{\beta}^{*},\bm{P}^{*}),\label{B2}
		\end{equation}
		which leads to a contradiction between~(\ref{B1}) and~(\ref{B2}).
		Accordingly, the assumption of existing another optimal solution of $(\bar{\bm{\alpha}},\bar{\bm{\beta}},\bar{\bm{P}})$ cannot hold, and thus $\left(\bm{\alpha}^{*},\bm{\beta}^{*},\bm{P}^{*}\right)$ is exactly the optimal solution to problem~(\ref{Dual}).
		
	\end{appendices}

	\bibliographystyle{IEEEtran}
	\bibliography{main}
\end{document}